# pVACtools v6: A comprehensive suite for neoantigen prediction, visualization, and therapy design

My H. Hoang*[,1], Susanna Kiwala*[,1], Megan Richters[1], Luke Hendrickson[1], Huiming Xia[1], Evelyn Schmidt[1], Christopher A. Miller[1,2], Kelsy C. Cotto[1], Cheryl F. Lichti[3], Zach Skidmore[1], Kartik Singhal[1], Jennie X. Yao[1], Jinglun Li[1], Mariam Khanfar[1], Isabel Risch[1], Jeffrey Ward[1,2], Ramaswamy Govindan[1,2], Gavin Dunn[1,2], Tanner M Johanns[1,2], S. Peter Goedegebuure[2, 4], Todd A Fehniger[1], Rachel Karchin[5,6,7,8], Robert D. Schreiber[2,3], William Gillanders[2,4], Obi L. Griffith[+,1,2,9,10], Malachi Griffith[+,1,2,9,10]

* These authors contributed equally; + Corresponding.

Affiliations
1. Division of Oncology, Department of Medicine, Washington University School of Medicine, St Louis, MO, USA
2. Siteman Cancer Center, Washington University School of Medicine, St Louis, MO, USA
3. Department of Pathology & Immunology, Washington University School of Medicine, St Louis, MO, USA
4. Department of Surgery, Washington University School of Medicine, St Louis, MO, USA
5. Department of Biomedical Engineering, Johns Hopkins University, Baltimore, MD, USA
6. Institute for Computational Medicine, Johns Hopkins University, Baltimore, MD, USA
7. Department of Oncology, Johns Hopkins University School of Medicine, Baltimore, MD, USA
8. Department of Computer Science, Johns Hopkins University, Baltimore, MD, USA9. Department of Genetics, Washington University School of Medicine, St Louis, MO, USA
10. McDonnell Genome Institute, Washington University School of Medicine, St Louis, MO, USA

## Abstract

With the rise of checkpoint blockade therapies and neoantigen-based vaccines reaching later-stage trials, there is a growing need for computational tools to identify and prioritize neoantigens. pVACtools, initially introduced in 2016[1], is an open-source informatic suite designed to support basic and translational neoantigen research. pVACtools assists prediction, prioritization, and visualization of neoantigens, as well as design of neoantigen-based therapies.

We describe several major advances to pVACtools since the last update: (1) expanded neoantigen quality and safety assessment features, including support for peptide presentation scoring, immunogenicity prediction, anchor residue analysis, reference proteome similarity, percentile score calculation; (2) addition of pVACsplice, a new tool for predicting neoantigens from tumor-specific cis-splicing mutations; (3) addition of pVACbind, a flexible tool that supports noncanonical neoantigen sources; (4) improvement in neoantigen selection strategies; (5) a substantially improved pVACvector algorithm that achieves higher DNA/mRNA vector vaccine design success rates with shorter runtimes; (6) new utilities to support synthetic long peptide vaccine design; (7) extended prediction support for many non-human species; and (8) addition of pVACcompare, a tool to support comparison between two pVACseq results. Together, these updates reinforce pVACtools as the field's most comprehensive toolkit for neoantigen research, from basic discovery to the design and execution of personalized cancer vaccine clinical trials.

# Introduction

Neoantigens, or tumor-specific antigens (TSAs), are peptides unique to tumor cells. Neoantigens may arise from multiple sources including somatic mutations, aberrant splicing events, endogenous retro-viruses and post-translational modifications. Neoantigen-MHC-TCR interactions are relevant to multiple therapeutic modalities, including (1) adoptive cell therapies (TCR-engineered T-cell therapy (TCR-T), Tumor Infiltrating lymphocyte therapy (TIL)), (2) antibodies (TCR-based T cell engagers, such as bispecific antibodies orTCR mimics), and (3) cancer vaccines (RNA vaccine, DNA vaccine, peptide vaccine, dendritic cell vaccine)[2]. Neoantigen-based therapies have been extensively evaluated in multiple clinical trials, either as standalone treatments or in combination with radiation, chemotherapy, targeted therapy, or other immune therapies. Early proof-of-concept for neoantigen therapy was established in melanoma through personalized dendritic[3] and long peptide vaccines[4,5]. These efforts have since matured into larger randomized trials: the phase 2b KEYNOTE-942 trial demonstrated that the mRNA-based neoantigen vaccine mRNA-4157 (V940), combined with checkpoint inhibitor pembrolizumab, significantly improved recurrence-free survival in resected high-risk melanoma compared to pembrolizumab alone[6]. At least three phase 3 trials are now underway (INTerpath-001, INTerpath-002, and INTerpath-009). Neoantigen-reactive tumor-infiltrating lymphocyte (TIL) therapies, including the FDA-approved lifileucel and NCI-led neoantigen-selected TIL infusion trials, have also demonstrated clinical responses in melanoma[7] and other solid tumors[8,9]. These results have catalyzed broader trials across tumor types and have placed the computational identification and prioritization of neoantigens at the critical interface between genomics and clinical care.

Prioritization of neoantigen candidates for therapy manufacturing, specifically vaccine manufacturing, is essential for various reasons. From a manufacturing standpoint, peptide synthesis is a cost-prohibitive and time-intensive process, making it infeasible to include more than a few dozen antigens per patient. For nucleic acid-based vectors, the package capacity of the construct imposes a hard limit: DNA, mRNA and viral vectors can only hold a certain number of bases before stability is compromised[10–12]. Similarly, the physical size of the AAV capsid limits the packaged DNA capacity[13]. Commonly, personalized cancer vaccines are manufactured with 20-30 epitopes. For example, the mRNA-4157 platform developed by Moderna can hold up to 34 epitopes[6], and the cevumeran platform developed by BioNTech is reported to encode 20 neoantigens[14]. Beyond logistical and technical constraints, there is immunological evidence that the immune system responds preferentially toward a subset of dominant epitopes, despite being exposed to numerous antigens - a phenomenon known as immunodominance[15–19]. These considerations collectively underscore the significance of computational tools that can guide candidate selection in an evidence-based manner.

Several computational frameworks have been developed to address aspects of neoantigen identification and prioritization. Tools such as Neofox[20], pTuneos[21], and LENS[22] provide algorithms for one or more steps in the neoantigen prediction and assessment workflow. However, many of these tools address individual components of the pipeline in isolation - for example, binding affinity prediction or immunogenicity scoring - without supporting the end-to-end workflow from antigen prediction to therapy design. These tools also vary in the genomic sources of neoantigens they support, the prediction algorithms implemented, and generation of reports and visualizations to facilitate design of personalized therapies. There remains a need for a comprehensive, modular, and openly available computational suite that supports the entire neoantigen workflow: from multi-source variant prediction and multi-dimensional quality assessment, through candidate visualization and selection, to downstream assistance with vaccine manufacturing. pVACtools was designed to fill this gap.

The first tool of the pVACtools suite - pVACseq[23] - was introduced in 2016, with the full pVACtools suite released in 2018[1]. It has since become one of the most widely adopted neoantigen analysis platforms in the field, with over 500 citations and approximately 225,000 downloads of the Python package and Docker images to date. The tools have been applied across a broad range of research and clinical contexts, including studies of immune evasion[24,25], tumor microenvironment evolution[26], the relationship between neoantigen burden and prognosis, mechanisms of response to immune checkpoint blockade[27–30], and the neoantigen landscape of numerous cancer types[31–34]. pVACtools has also directly supported the development of mRNA-based, DNA-based, and peptide cancer vaccines and has been used in more than 12 neoantigen-related clinical trials at Washington University alone (NCT03422094, NCT03606967, NCT03199040, NCT05111353, NCT03532217, NCT02348320, NCT04397003, NCT03122106, NCT03988283, NCT03956056, NCT05741242, NCT03121677). With this widespread adoption and the rapid expansion of neoantigen-based therapies entering clinical development comes a need to expand the suite's features to enable the growing user community to benefit directly from new scientific advances within a familiar, integrated workflow.

Here, we describe the novel features and tools added to pVACtools since its introduction. These updates facilitate neoantigen prediction, prioritization, safety evaluation, and streamline the process of neoantigen therapy production.

# Results

pVACtools is an open-source suite of modular Python and R-based components that integrates with DNA/RNA sequencing pipelines[35] and is designed to support neoantigen prediction, candidate selection, and therapy manufacturing (**Figure 1, Figure 2**). The suite originally comprised: pVACseq for SNV/indels, pVACfuse for fusion neoantigen prediction, and pVACvector for DNA vaccine vector design. Since its introduction, we have also added pVACsplice, for splicing neoantigen prediction; pVACbind, for custom or non-canonical peptides binding prediction; pVACview[36] , for interactive visualization and candidate review; and pVACcompare, for cross-run comparisons. pVACtools is also adopted in ImmunoNX[35], an end-to-end WDL workflow which takes in raw sequencing data and performs variant calling, neoantigen prediction and prioritization for therapy production.

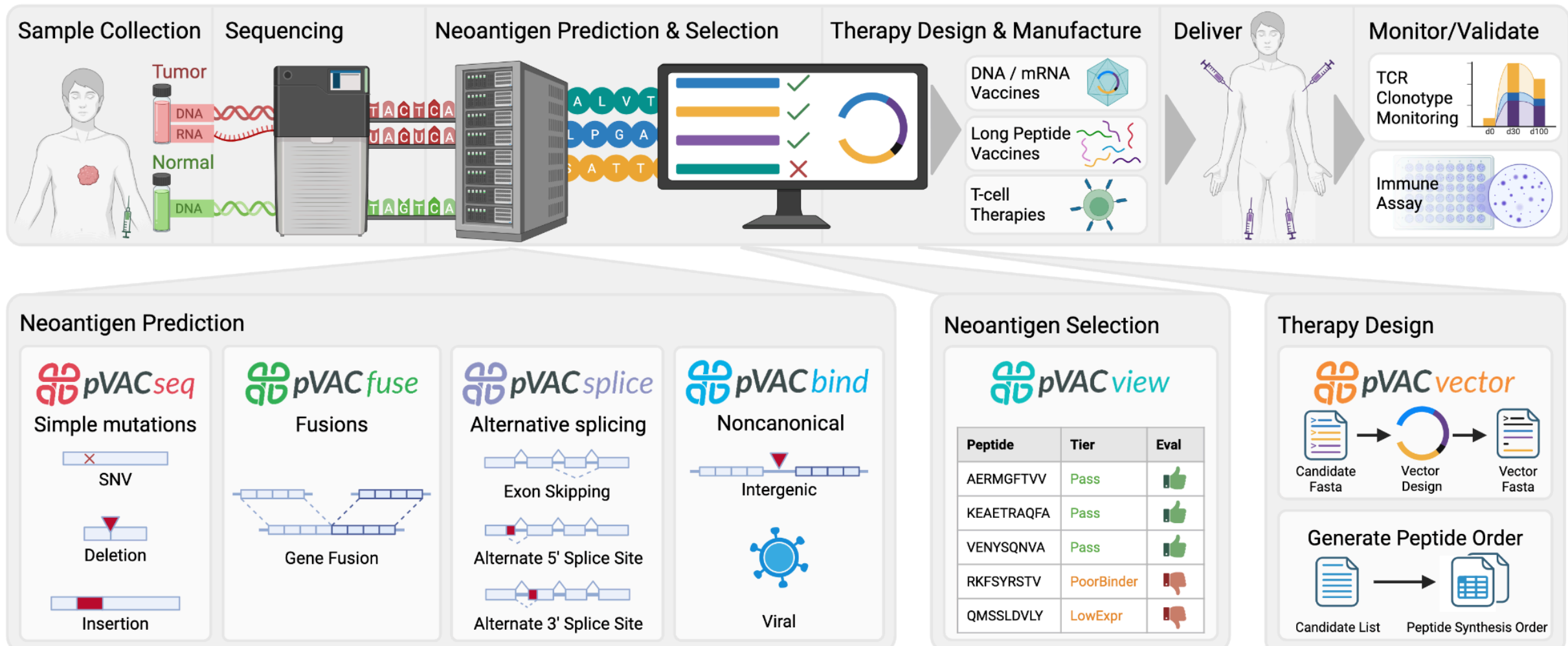

**Figure 1. pVACtools aids users in prediction, selection, and design of neoantigen candidates for anti-cancer therapies.** The workflow to create neoantigen therapies starts with collection of a tumor biopsy and matched normal sample, followed by genomic sequencing, neoantigen prediction, selection of neoantigen candidates, design, manufacturing, delivery of therapy to the patient, and immune monitoring post-therapy (for example, ELISpot to confirm pMHC-TCR induced response, or scTCRseq to track potential T-cell clonal expansion following therapy). Within this process, pVACtools offers a set of bioinformatic tools that facilitate all steps from prediction through therapy design. Specifically, pVACseq, pVACfuse, and pVACsplice predict and prioritize neoantigen candidates from SNVs/indels, gene fusions, and cis-splicing variants, respectively. pVACbind predicts neoantigen characteristics from user-input peptides, assisting non-canonical neoantigen prediction. pVACview presents a platform to visualize these candidates, aiding candidate selection for therapies. pVACvector assists the design of DNA vaccine vectors. pVACseq also offers sub-commands that facilitate ordering of long peptide vaccines for manufacturing.

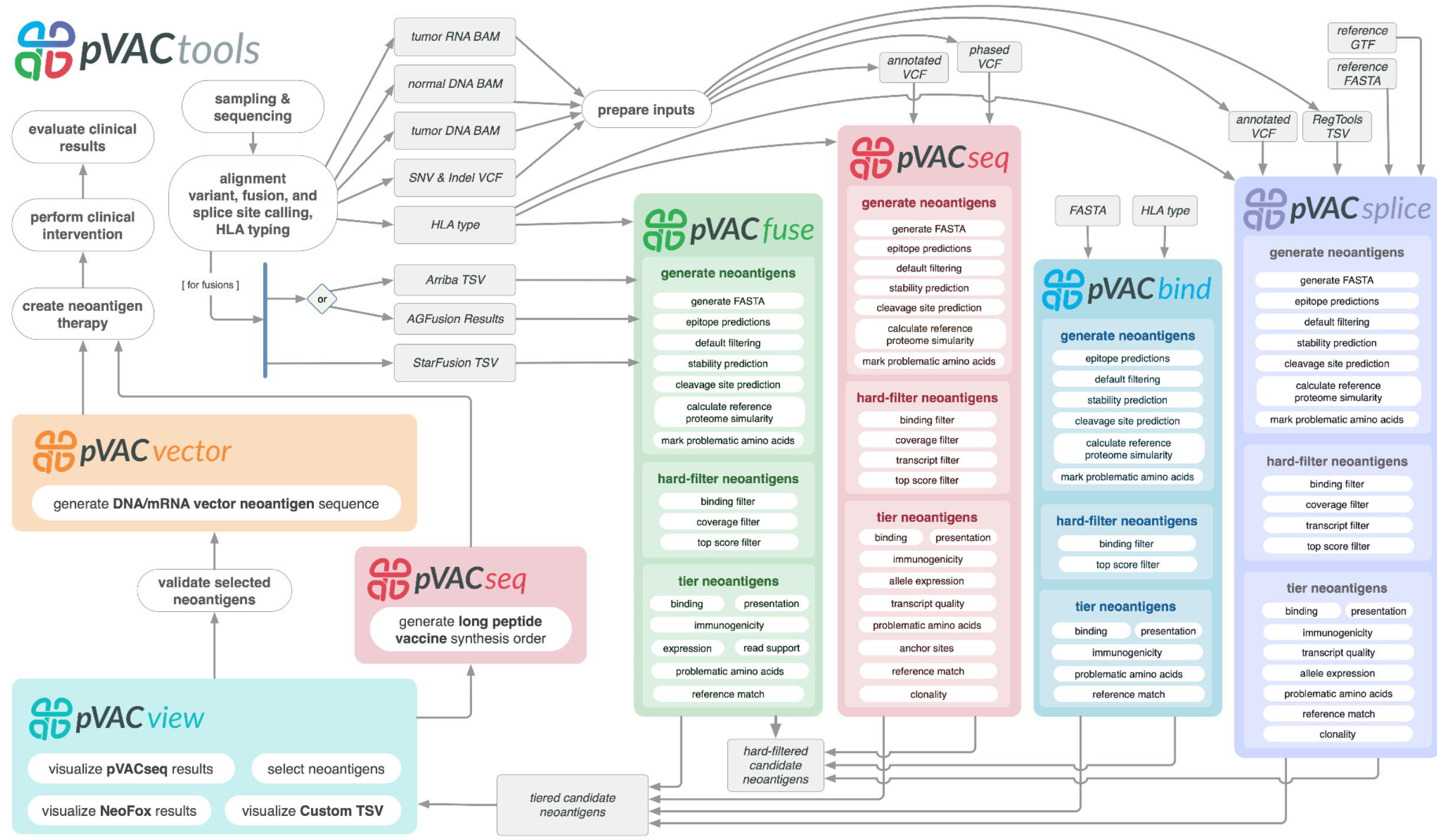

**Figure 2**. **Summary of main modules in the pVACtools suite, inputs and output features**. An upstream pipeline (e.g., ImmunoNX) takes raw sequence data and performs alignment, variant calling, and HLA typing). These pipeline results are the inputs for prediction modules (pVACseq, pVACfuse, pVACsplice, pVACbind) and include variant source (annotated VCF, fusion call TSV, novel junction TSV, peptide fasta), HLA type and reference files. The prediction modules are executed as a set of steps that include neoantigen generation, filtering and tiering. Neoantigen generation involves epitope prediction, stability prediction, cleavage site prediction, proteome similarity measures, etc. Filtering considers binding predictions, coverage, top score, etc. Tiering considers binding, presentation, immunogenicity, reference matches, expression, and more depending on the module. Candidates can be visualized with pVACview. Users can generate mutated peptide candidates of desired length with the 'generate protein fasta' command available in each pVACtools prediction module. Taken one step further, one can create long peptide vaccine synthesis order forms directly from a subcommand available in the pVACseq prediction module. Alternatively, users can generate multi-epitope DNA/mRNA-vector sequences using pVACvector by supplying a list of candidate peptides as input.

# 1. Improved neoantigen quality assessment

In the original paper[1] pVACtools offered several features to evaluate neoantigen characteristics, including peptide(neoantigen)-MHC binding affinity, peptide processing potential, peptide-MHC stability, DNA allele fraction, RNA expression, MHC peptide binding promiscuity (capability to bind to multiple MHC alleles). Since then, we've added a host of novel features to improve neoantigen quality assessment in 3 major areas: peptide presentation, dissimilarity to self, and immunogenicity. We have also added percentile scores to allow more robust comparison between algorithms. A summary of all neoantigen quality assessment criteria (old and new) available for neoantigen prediction tools (pVACseq, pVACbind, pVACfuse, pVACsplice) are presented in **Figure 3**. Details of novel features are as follows:

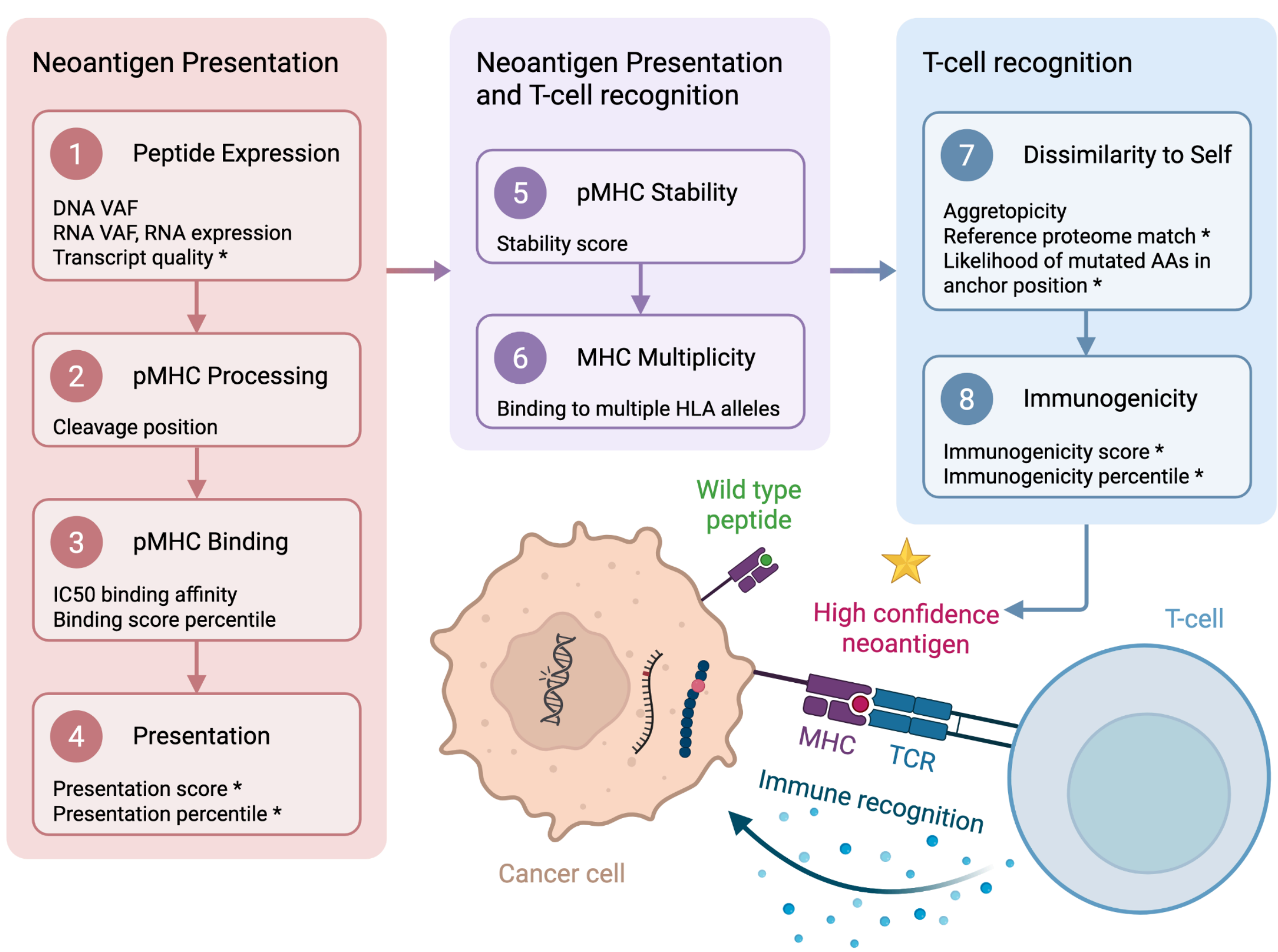

**Figure 3. pVACtools features that facilitate neoantigen prioritization**
pVACtools has multiple features to interrogate peptide binding to MHC complexes and T cell recognition of neoantigen candidates. In the original pVACtools publication we described features supporting assessment of peptide expression (through VAF, expression, transcript support level), peptide-MHC processing probability (through cleavage position prediction with NetChop), peptide-MHC binding (through IC50 binding affinity score from 11 algorithms), peptide-MHC stability (through stability prediction with NetMHCstabpan), and ability of peptide to bind to multiple MHC alleles. Since that publication, pVACtools developments have focused on adding support for modeling MHC presentation probability (predicted by 4 algorithms), multiple criteria assessing dissimilarity of mutant peptide to self, and immunogenicity probability (predicted by 2 algorithms). We've also added MANE Select and Canonical status as additional metrics for evaluating transcript quality. New features since initial publication are denoted with an asterisk (*).

## 1.1. Peptide presentation scores predict likelihood of peptide-MHC to be presented on tumor cell surface

Peptide presentation scoring provides an integrated measure of the likelihood that a peptide will be processed and presented on the MHC complex. Presentation prediction models are trained on mass spectrometry data from peptides experimentally isolated from cell-surface MHC molecules (also known as immunopeptidomics data)[37]. Therefore, unlike binding affinity scores, which primarily reflect the direct interaction between peptide and MHC, presentation scores capture the combined effect of binding affinity, peptide-MHC complex stability, and sequence-based determinants of antigen processing such as proteasomal cleavage patterns.

Since version 1.5, we have added support for presentation scoring for MHC class I via 3 algorithms (MHCflurryEL[38], NetMHCpanEL[39], BigMHC-EL[40]), and for MHC class II via NetMHCIIpanEL[39] .

## 1.2. Improved safety via self-similarity checks

All T cells undergo negative selection, a process that eliminates autoreactive T cells, preventing them from attacking the body's own tissues. Therefore, neoantigens bearing greater resemblance to self-peptides are unlikely to elicit T cell responses. pVACtools provides three approaches to assess neoantigen dissimilarity to self, as a safety assessment feature. The original pVACtools release introduced Agretopicity, the ratio of wildtype to mutant IC50 binding scores, reflecting binding strength of mutant peptide to MHC relative to its wildtype counterpart. Two additional features have since been added: reference proteome similarity comparison and anchor residue prediction which enhances interpretation of agretopicity values.

Users can check for similarity between predicted epitopes and reference proteome for all neoantigen prediction tools (pVACseq, pVACbind, pVACfuse, pVACsplice). This step searches predicted epitopes against the human reference proteome, flagging any candidate with significant similarity to a known self-peptide. Users can use BLASTp to search against either the *RefSeq Select* protein database (default) or the full *RefSeq* protein database. As a more performant alternative, a k-mer string search against a user-provided custom peptide fasta may be performed. When running the search with BLASTp, users have the option of using the NCBI Protein BLAST API (default) or a standalone BLAST installation.
Anchor residue prediction provides a complementary and more structurally-informed dissimilarity assessment. Anchor residues are amino acid positions within a peptide that make direct contact with the MHC binding groove pockets, while non-anchor residues are more exposed toward the TCR. When a mutation falls at a non-anchor position, the TCR-facing residues of the mutant and wildtype peptides differ, enabling T cell discrimination and potential immune activation despite agretopicity status. Conversely, when the mutation falls at an anchor position, the TCR-facing residues of both peptides are identical despite their different MHC binding behaviors, making productive T cell engagement unlikely in case of low agretopicity. pVACseq incorporates prediction of anchor residues for MHC class I alleles trained by a machine learning model[41], and also assigns a tier 'Anchor' to any neoantigen with mutated amino acid at a predicted anchor residue.

### 1.3. Immunogenicity scores predict T-cell engagement

Not all peptides with good binding affinity that are successfully presented by the MHC on the cell surface are necessarily recognized by T-cells. There have been multiple efforts to develop algorithms estimating immunogenicity - the probability that a given pMHC complex will elicit a T-cell response - by training models on experimental T-cell activation assays. To support this concept we have added support for two MHC Class I immunogenicity prediction algorithms: BigMHC_IM[40] and DeepImmuno[42]. BigMHC_IM uses BigMHC_EL - a model trained on mass spectrometry eluted ligand data - as a base model, then applies transfer learning on immune assay data, to predict which presented peptides can actually trigger T-cell expansion. DeepImmuno-CNN predicts the immunogenicity of MHC-peptide pairs by using a convolutional neural network to score the pMHC complex based on T-cell assays in IEDB and amino acid physicochemical features. Together, these algorithms enable users to prioritize candidates not just for strong MHC binding, but for functional immunogenic potential. Together, these algorithms enable users to prioritize candidates not just for strong MHC binding, but for functional immunogenic potential.

### 1.4. Percentile scores enable cross-algorithm comparisons

Along with raw peptide-MHC binding affinity score, presentation score, and immunogenicity score, pVACtools also provides corresponding percentile scores from selected algorithms. A percentile score represents the rank of a given peptide's predicted (binding affinity/presentation/immunogenicity) score relative to a large reference set of random peptides scored by the same algorithm, expressed as a percentage. A lower percentile score indicates strong predicted binding/presentation/immunogenicity, as it means fewer random peptides score better than the peptide of interest. Percentile scores enable meaningful cross-algorithm comparison by normalizing for differences in the absolute score scales used by different prediction tools. Users may optionally use these percentile scores for filtering and tiering. Additionally, users are able to control whether both or either the binding and the percentile thresholds need to be met in order for a neoantigen candidate to be considered.

## 2. More robust fusion detection and prioritization in pVACfuse

pVACfuse provides neoantigen predictions from gene fusion events, but has previously been limited to results compatible with AGfusion. In this release, we broadened its reach by adding support for the popular Arriba[43] package (**Figure 2**). In its original implementation, pVACfuse was also unable to consider key RNAseq metrics for prioritization, but with the addition of STAR-fusion[44] support, it can now access information on overall gene expression and read support. These details are then propagated downstream, allowing fusion-derived neoantigens to be filtered, prioritized, and tiered in a more robust manner.

# 3. Support for splicing-induced neoantigens: pVACsplice

Aberrant mRNA splicing is a common feature of tumor transcriptomes, generating tumor-specific transcript isoforms that can give rise to neoantigens absent from normal tissues. Splicing neoantigens have historically been underrepresented in neoantigen discovery pipelines, in part due to the computational complexity of identifying tumor-specific junctions and translating them into candidate epitopes. To address this gap, we introduce pVACsplice: a tool that predicts neoantigens from tumor-specific alternative splicing junctions identified from paired whole-exome and RNA sequencing data.

## Tool inputs and usage

pVACsplice takes four primary inputs: (1) a RegTools[45] TSV output file containing candidate splicing junctions and their associated somatic variants generated with 'regtools cis-splice-effects identify'; (2) a VEP-annotated somatic variants VCF file; (3) a reference genome FASTA file; and (4) a reference transcript annotation GTF file. As with other pVACtools prediction tools, users also specify HLA alleles, desired epitope lengths, and preferred neoantigen prediction algorithms. It is important that the same reference genome and annotation files are used across the variant calling, junction identification with RegTools, and pVACsplice run steps, to ensure accurate transcript reconstruction (**Figure 2**).

## Tool architecture and outputs

pVACsplice processes inputs through three sequential steps: junction filtering, alternative transcript annotation (predicting the impact of alternative junction usage on protein sequence), and neoantigen extraction and prioritization (**Figure 4**).

To focus prediction on high-confidence, protein-coding splicing events, pVACsplice applies four default filters to the RegTools junction output. First, junctions are required to meet a minimum read support threshold (default: 10 reads) to exclude lowly-expressed events. Second, junctions are filtered by splice site category: by default, only junctions involving at least one known splice site (novel acceptor, donor or exon-skipping) are retained, as these carry additional evidence for biological authenticity. Third, only junctions associated with protein-coding transcripts and predicted to impact their coding sequence, as defined by the reference GTF biotype annotation, are carried forward. Finally, variants must fall within a defined distance (default: 100 bp) of a junction edge to be considered likely cis-regulatory contributors to the observed splicing event. All filter thresholds are user-configurable to be more or less stringent.

For each passing junction, pVACsplice constructs an alternative transcript by replacing wild-type splice coordinates with the novel junction coordinates identified by RegTools. Alternative 3'/5' splice site events are handled by identifying the nearest reference exon boundary and adjusting the coordinate accordingly; exon skipping events are modeled by removing the intervening exons from the coding sequence. The resulting DNA sequence is translated from the canonical start codon using the reference FASTA, with strand orientation accounted for. Intron retention and mutually exclusive exon events are not currently supported. Importantly, the relationships between junctions, transcripts, and variants are not one-to-one: a single junction may be associated with multiple reference transcripts, multiple variants may contribute to the same junction, and multiple splice junctions may be associated with a single variant, all of which are reported to allow downstream prioritization.

Next, wild-type and alternative transcript pairs are compared at the k-mer level for each desired epitope length. Peptide sequences unique to the alternative transcript are extracted and submitted to the pVACtools neoantigen prediction pipeline, which scores each candidate for MHC binding affinity, presentation, and immunogenicity using user-specified algorithms as described above.

pVACsplice main outputs include a set of epitope files reporting all candidates, candidates meeting filtering threshold, and top ranked candidate resulting from each novel junction. pVACsplice also provides additional outputs including: 1) a GTF containing RegTools-identified novel junctions 2) the coordinates of novel junctions and their associated cis-splicing mutations 3) the full sequences of the wild-type and splice-altered transcript and protein.

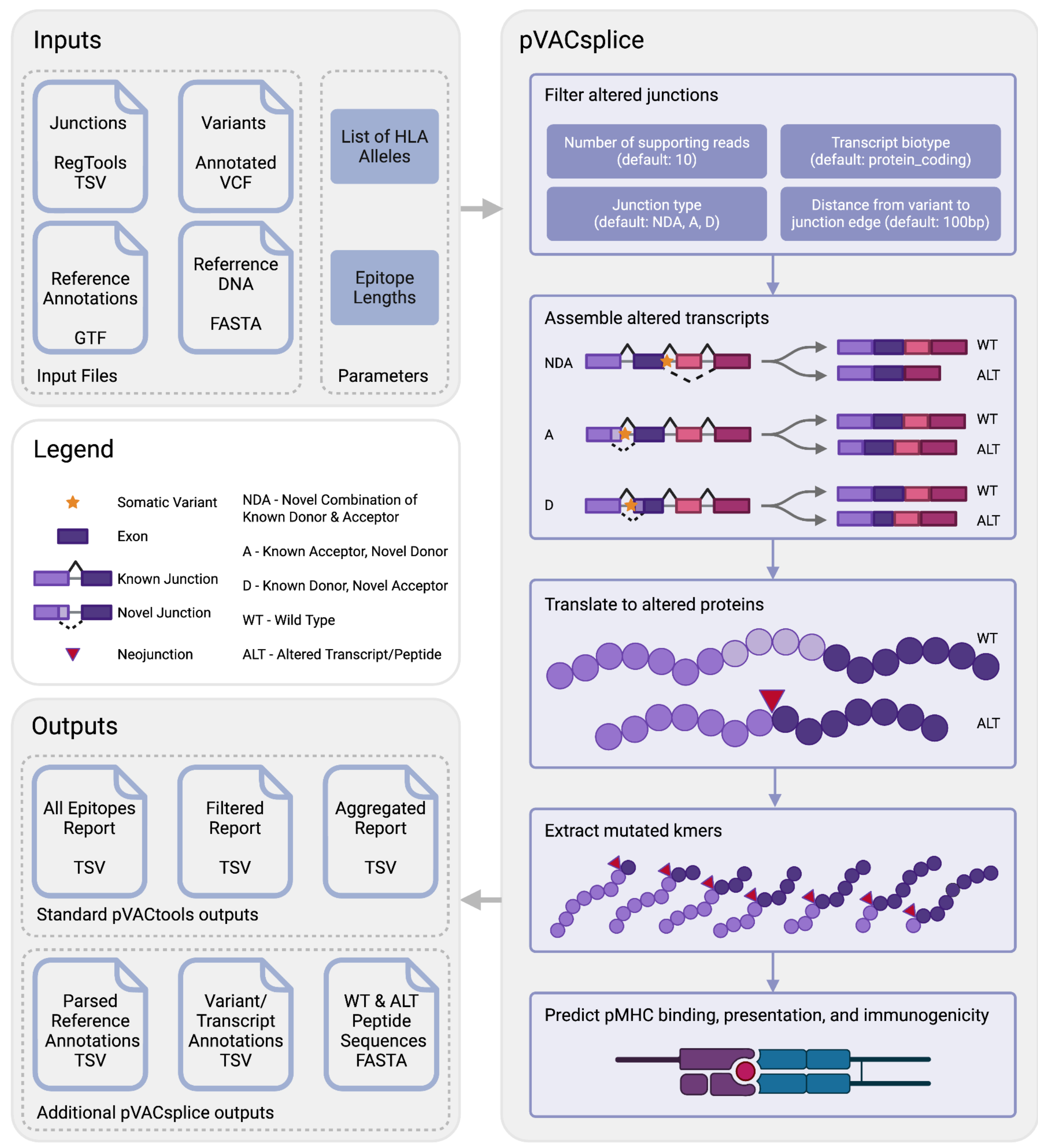

**Figure 4**. **pVACsplice assists prediction of neoantigens from cis-splicing mutations**. After reading key inputs (junctions, variants, and other files/parameters), pVACsplice applies filters to produce a short list of high quality cis-splicing associated junctions with protein coding potential. It then generates altered transcript models and translates them into full-length altered protein sequences. Next, all novel k-mers (short peptides containing the mutated amino acids), are extracted and analyzed using the multiple prediction algorithms in the pVACtools suite.

# 4. Support for non-canonical antigens: pVACbind

Beside the tools specializing in predicting and prioritizing neoantigen from certain somatic variation sources (pVACseq, pVACfuse, pVACsplice), we also offer pVACbind, a tool that predicts antigen properties from peptide sequences supplied in fasta format (**Figure 2**). Given the input sequence peptide, pVACbind generates overlapping k-mer windows and predicts binding affinity, presentation, immunogenicity, and similarity to reference proteome for these k-mers. Given its simplicity and flexibility, pVACbind allows exploration of novel sources of antigens which haven't yet been streamlined into pVACtools with dedicated modules (eg. antigens from intergenic regions, viral sources, etc.).

# 5. Improved Neoantigen Selection

One of the key functions of pVACtools after neoantigen prediction is the evaluation of each candidate by selecting the most promising neoantigen for each variant, assigning them a tier that reflects each top candidate's suitability for inclusion in the final immunotherapy design, and providing GUI capabilities for final evaluation and selection of candidates. This process has been extensively revised and expanded since the initial pVACtools publication as follows.

## 5.1 Tiering of candidates allows for easy candidate ranking

To facilitate neoantigen review, the primary report (previously condensed output file, now aggregated output file) has been revamped to more clearly represent the best neoantigen candidate for each variant, and assigns it to a tier. These tiers encapsulate a number of factors that reflect suitability for inclusion in personalized immunotherapy designs, validation experiments and other downstream applications (**Figure 2**).

Selection of the top neoantigen candidate for each variant takes into consideration various factors. Candidates with a low IC50 binding affinity value are prioritized, while candidates failing the anchor assessment or containing user-defined amino acids problematic for vaccine manufacturing are deprioritized. In the case where different transcript isoforms result in separate sets of candidates for the same variant, candidates resulting from well-supported protein-coding transcripts are prioritized. Transcript quality is assessed using Ensembl transcript support level (TSL), MANE Select status, and canonical transcript designation.

The neoantigen candidate Tier assessment takes into account many of the criteria described in Section 1, including binding affinity, peptide processing, expression, coverage, anchor residue status, and reference proteome similarity (**Supplementary Table 1**). The available tiers depend on the specific pVAC prediction tool but generally include Pass, Poor Binder, Reference Match (Ref Match), and Poor tiers. If users specify a list of problematic amino acids in the prediction step, tier Problematic Position (Prob Pos) will be available. Tools like pVACseq and pVACsplice that include expression and coverage information will also have tiers for Subclonal, Low Expression (Low Expr), No Expression (No Expr) and Poor Transcript variants while pVACfuse supports Low Read Support and Low Expression (Low Expr) as additional tiers. Additionally, an Anchor tier bins variants where the best peptide fails the anchor assessment in pVACseq.

### 5.2. The pVACview user interface supports comprehensive candidate evaluation and selection

Starting from version 3, we retired pVACapi and pVACviz, and introduced pVACview[36] as the tool for visualization of neoantigen prediction results. pVACview works with the pVACseq aggregated report described above (**Supplementary Figure 1**) and provides a wide range of additional visualizations to further illustrate the variables that need to be considered when selecting neoantigen candidates for inclusion in an immunotherapy. For example, per-algorithm binding affinity IC50 and percentile scores are visualized in violin plots to check concordance between algorithms. Anchor sites are illustrated in heatmaps to facilitate evaluation. Reference proteome matches are visualized in the context of the larger protein sequence to assess their potential impact. Additionally, details about alternate neoantigen candidates resulting from each variant can be explored alongside the best neoantigen candidate originally identified. Users can load both MHC class I as the primary set of candidates and MHC class II candidates as additional information, or vice versa. Outputs from other neoantigen prediction pipelines (Neofox, etc.) can also be visualized.

## 6. Improved and expanded immunotherapy design functionalities

Cancer vaccine design requires careful consideration of safety and practical manufacturing constraints. Here, we describe the refactoring of pVACvector to optimize multi-epitope vector-based vaccine design, and introduce new functionalities to aid synthetic long peptide vaccine design.

### 6.1 Refactor of pVACvector for efficient DNA vaccine design

pVACvector is designed to aid in the construction of multi-epitope DNA/mRNA/viral vector based personalized cancer vaccines. It attempts to find an optimal ordering of input peptide sequences that avoids strong binding junctional peptides (i.e. those formed at the boundary between adjacent input sequences). To find this ordering, pVACvector constructs a directed graph, in which each node represents a peptide sequence, and each directed edge represents the junction formed by placing one peptide immediately after another. Edges are only added for “valid” junctions where none of the junctional peptides are well-binding. After constructing the graph, a simulated annealing procedure is used to find a path through the graph (i.e., a possible ordering of the peptides), that maximizes the combined binding affinity of the junctional epitopes across the full sequence.

pVACvector has undergone significant improvements since it was last described[1]. The initial algorithm had two key limitations in its handling of invalid junctions. First, when spacer or clipping operations were used to resolve an invalid junction, the graph was rebuilt from scratch, discarding valid junctions that had already been identified in earlier iterations. Second, the clipping procedure tested only one of the three possible combinations for each invalid junction (clipping the C-terminus of peptide A and the N-terminus of peptide B simultaneously), while the two single-peptide clipping combinations (clip A only; clip B only) were not explored. These limitations were addressed in version 5: the graph is now built iteratively, retaining previously identified valid junctions across spacer and clipping attempts, and all permutations of clipping combinations are tested for each invalid junction. Additionally, as the input fasta now includes the core neoantigen sequence, pVACvector can avoid clipping amino acids in the core sequence, preserving the functionally critical region, especially when the binding core candidate is situated near the terminus of a subunit candidate sequence. Finally, pVACvector can now identify partial solutions by excluding peptides if a solution with all input peptides can't be found. (**Figure 5**).

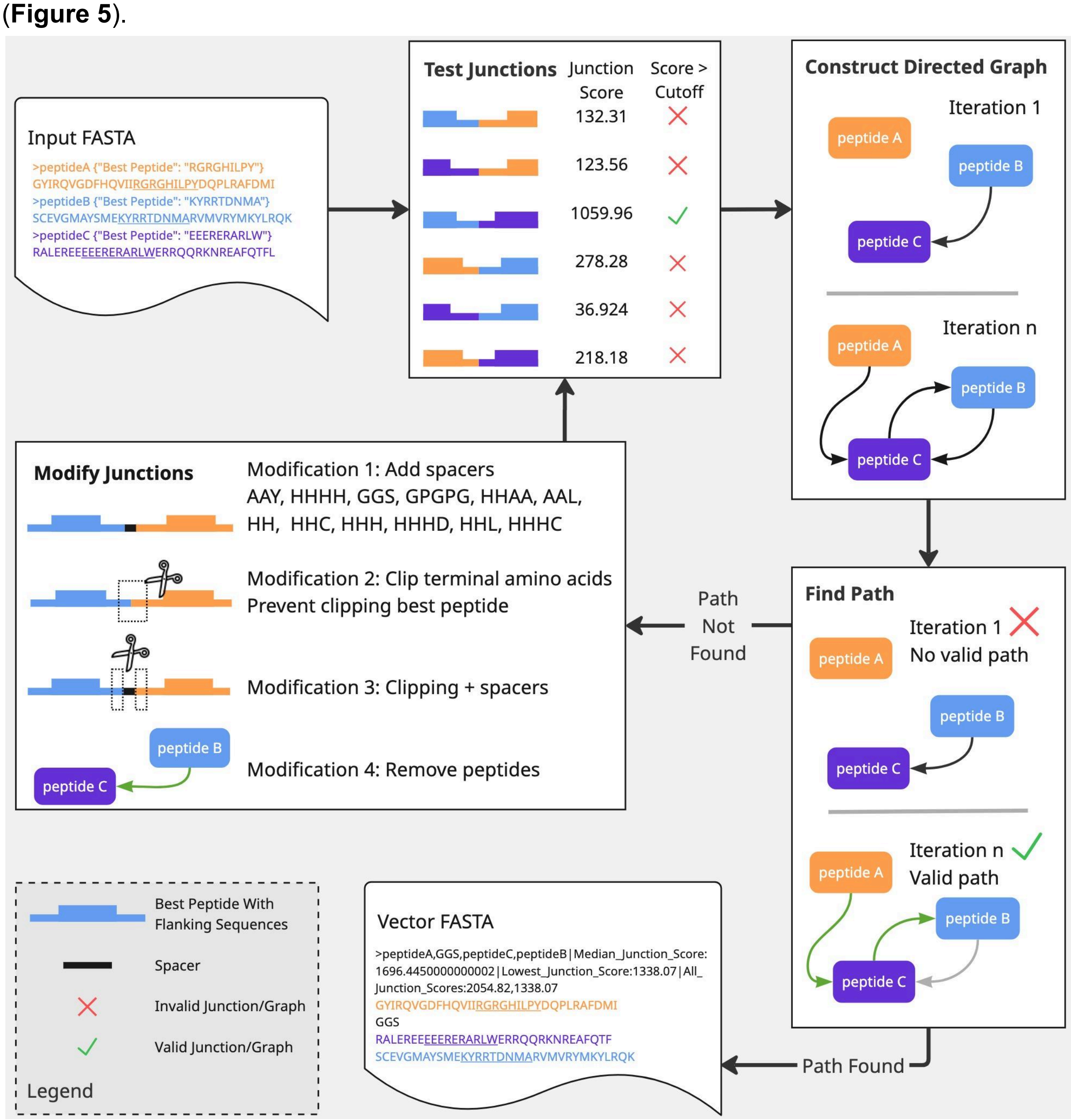

**Figure 5**. **Main operations in pVACvector vector construction in version 5+**. User-provided subunit candidate sequences are first joined pairwise and the junctions examined for well-binding kmer peptides. Valid junctions without any well-binding kmers will be added to the graph as a directional edge between the two peptide nodes. A simulated annealing procedure is run to find a path through the graph that includes all peptide nodes. If no valid solution is found, modifications to the junctions will be tested iteratively. First, spacers (user-specified linkers) will be added between tested junctions. Second, peptides will be clipped at each end (or both ends) of tested junctions. Third, spacers and clipping will be done in combination. Lastly, peptides can be excluded to explore partial solutions. Finding a valid path at any point of the process will result in pVACvector succeeding and creating a Vector FASTA file of the final peptide ordering including spacers and clipped peptides and excluding removed peptides where those modifications were necessary for the solution.

## 6.2 Functionalities for evaluating peptide manufacturability and designing synthetic long peptide vaccines

### 6.2.1 Manufacturability metrics allow identification of difficult to synthesize neoantigens

We incorporate eight manufacturability criteria proposed by VaxRank[46] to assist users in identifying peptide candidates that may be difficult to synthesize. These criteria are reported in the manufacturability TSV output of every pVACtools prediction tool. The criteria address five categories of synthesis concern. First is cysteine content (examined by cysteine count metric and a C-terminal cysteine flag), as cysteine residues can form unwanted disulfide bonds that interfere with synthesis and purification, and cysteines can undergo oxidation which affect stability of vaccine during storage[47,48]. Second is proline placement (a C-terminal proline flag), as proline introduces difficulties in chain elongation in Solid-Phase Peptide Synthesis (SPPS), reduces yield and introduces impurities[49] in vaccine synthesis process. Third is asparagine context  (N-terminal asparagine and asparagine-proline bond count), as N-terminal asparagine is difficult to synthesize (asparagine N-terminal protecting group can be difficult to remove during cleavage)[50], and asparagine-proline bonds are susceptible to cleavage, leading to potential peptide fragmentation at those sites, affecting vaccine stability[51]. Fourth is hydrophobicity (presented by the Grand Average of Hydropathicity (GRAVY) score of the C-terminal 7-mer and the maximum 7-mer GRAVY score across the peptide) as long hydrophobic stretches increase the risk of aggregation and poor solubility, forming injection-site depots that interfere with T-cell response[52,53] . The last criterion is difficult N-terminus residual, which are flags for glutamine, glutamic acid, or cysteine - all of which present stability and storage challenges[54]. Together, these metrics enable users to deprioritize sequences with unfavorable physicochemical properties before committing to synthesis.

### 6.2.2 Novel functionalities for synthetic long peptide vaccine design

During the peptide vaccine design process, users might be interested in generating long peptide sequences (e.g. of sizes 20-30 amino acids) that capture multiple epitopes (for both class I and class II MHC) and improve uptake and processing in antigen processing cells. To this end, users can use the ‘generate_protein_fasta’ function in pVACseq, pVACfuse, and pVACsplice to create a list of mutated peptide sequences and wildtype peptide sequences, when available, with desired flanking sequence size. This command supports many of the same options available in the main pipelines, such as incorporating proximal variants, limiting long peptides to those resulting from PASS-only variants or with specific VEP-predicted biotypes, and excluding transcripts with incomplete 5’ or 3’ positions.

To further streamline candidate review and synthesis preparation, users can also run ‘pvacseq create_peptide_ordering_form’ to generate a list of peptide ordering files (**Figure 2**). These files include: (1) a fasta containing mutated peptide sequences; (2) a manufacturability report with metrics such as cysteine content, hydrophobicity, and sequence complexity for each peptide; (3) a visually annotated spreadsheet with long peptide sequences (highlighting mutated sequence, class I and II epitopes, and problematic residues), their respective molecular weights, and more; and (4) an Annotated neoantigen candidates excel sheet, which guides manual review of selected variants[35]. (**Supplementary Figure 2)**

# 7. Support for non-human species facilitates basic research and translational applications

Non-human animal models remain indispensable for studying tumor immunology and evaluating neoantigen therapy efficacy prior to clinical translation. Furthermore, a growing number of anti-cancer therapy trials are being conducted in veterinary species such as dogs and cats[55–58].

pVACtools extends neoantigen prediction support to 33 non-human species although not all pVACtools functionalities are available for all species. Anchor residue prediction, which informs neoantigen dissimilarity to self, is currently available only for human and mouse. For finding matches of neoantigen candidates against the reference proteome, the default `refseq_select_prot` BLASTp database only natively supports human and mouse species. However, users can select to use the `refseq_protein` BLASTp database with any of the supported species. Alternatively, users may provide a species-of-interest peptide fasta to run this feature without using BLASTp. For pMHC stability predictions, NetMHCstabpan only supports human, mouse, non-human primates (chimpanzee, macaque, and gorilla), pig, dog, and horse. Despite these limitations, the vast majority of pVACtools functionalities are available in all supported species, enabling robust non-human neoantigen research.

# 8. pVACcompare enables comparison of pVACseq outputs to validate results:

Systematic comparison of pVACseq runs is essential for validating result fidelity, optimizing pipeline parameters, supporting reproducibility, and ensuring the rigor required for clinical-grade applications. We introduce a new tool, pVACcompare, to assist these analyses. The tool can be used to compare outputs of two pVACseq runs from different pVACtools versions, runs with different parameters, or runs with different inputs (e.g., two different biopsies regions of the same tumor). The tool outputs comparisons of pVACseq runs inputs, all epitopes and aggregated epitopes files, reference matches and run metrics (binding threshold, epitope lengths, alleles, etc) (**Supplementary Figure 3**).

# 9. Data analysis with pVACtools

To demonstrate the performance of several of pVACtools' new features, we conducted a series of example analyses as follows..

## pVACsplice identified putative splice variant neoantigen across GBM and SCLC cohorts

To demonstrate the utility of pVACsplice in neoantigen exploration, we applied it to two independent tumor cohorts. The first comprised a glioblastoma (GBM) cohort (n=19 patients) and brain metastasis (BrMet) cohort (n=11 patients), analyzed together as a combined brain tumor cohort of 30 patients with 80 total samples, where each patient tumor had multiple biopsy sites[59]. Cis-splicing events were detected with RegTools[45] with default parameters (2 bp into the intron and 3 bp into the exon from each exon edge), and neoantigens for respective patient HLA types were predicted with pVACsplice. We identified 1,264 unique tumor-specific splice junctions absent from GTEx[60]. 11/30 patients (21/80 samples) had splicing variants with robustly supported events (alternative splicing junctions with at least 10 supporting RNA reads) and corresponding junctions absent from GTEx. 7/30 patients (11/80 samples) had at least 1 predicted strong-binding epitope (IC50 ≤ 500 nM) for the patient's HLA alleles. These findings are notable given GBM's characteristically low tumor mutation burden, where SNV-derived neoantigens are relatively sparse, splicing neoantigens may therefore represent a meaningful additional source of therapeutic targets in this tumor type. Some variants and neoantigens were shared across sites, but many were private, further highlighting GBM's intra-tumor spatial heterogeneity (**Figure 6**).

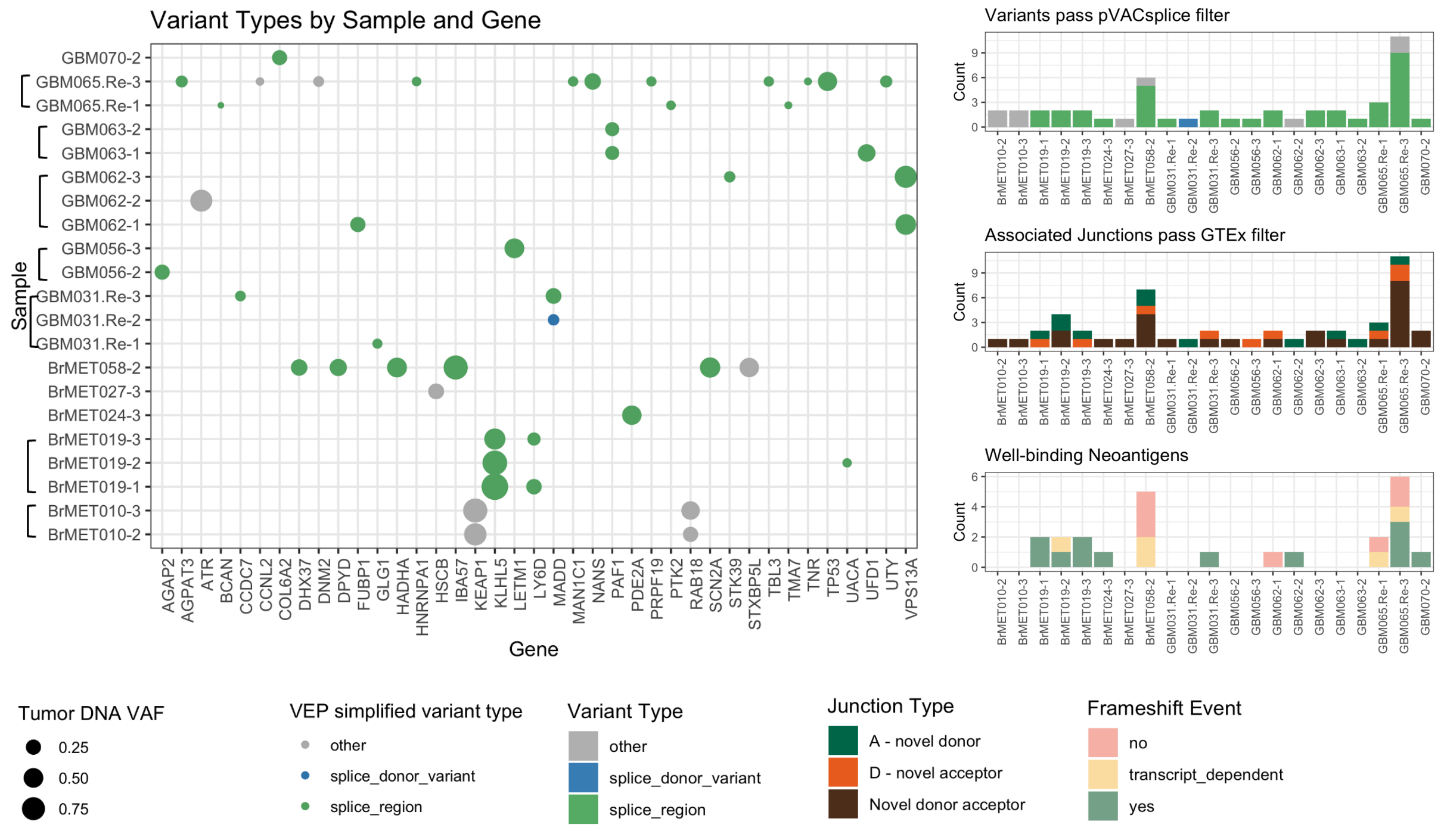

**Figure 6**. **pVACsplice predicts multiple cis-splicing variants, some of which are predicted to yield well-binding neoantigens in 21 GBM/Brain metastasis samples (11 patients)**. **Left panel**: a dot plot summarizes the cis-splicing variants observed across the cohort. Dots are sized according to variant tumor DNA VAF, and colored based on VEP annotation. As expected, some variants are shared across different biopsy regions, whereas some variants are private (specific to a specific biopsy region). **Right panel**: A summary of pVACsplice outputs including splice variants, junctions, and predicted MHC binders across the cohort. Variant counts in this panel refer to the number of variants passing the pVACsplice default filters. Junction counts illustrated here are for junctions associated with these variants that are confirmed as absent from normal tissues in GTEx. Neoantigen counts refer to the number of peptides from a unique gene with at least one well-binding epitope to the patient's HLA alleles. The Variant to Junction ratio is not always 1:1, as in some cases, a single cis splice variant can result in more than one alternate splicing event. Most but not all junctions associated with cis-splicing that pass pVACsplice filters are also absent from GTEx, and neoantigens can arise from peptides from either a frameshift or inframe event.

In a second application, pVACsplice was used to explore splicing neoantigens in a small cell lung cancer (SCLC) cohort (n=57 samples). We identified 6,154 unique tumor-specific splice junctions absent from normal tissues in GTEx. 42/57 samples had one or more splicing neoantigens from robustly supported splice events, 34/57 samples had strong-binding epitopes, including in key drivers TP53 and RB1 - the two most frequently altered tumor suppressor genes in SCLC (**Supplementary Figure 4, Supplementary Figure 5**).

## New pVACvector algorithm enables faster, higher success rate, and more optimized cancer vaccine vector designs

To assess efficiency of the updated pVACvector algorithm, we benchmarked pVACvector version 5 against version 4 across 11 vaccine design cases under 3 parameter settings: “Default”, “IC1500” and “IC1500_Clip5”. The default setting used a junctional binding affinity limit of 500 nM, and maximum clipped peptide length of 3 aa. The IC1500 setting used a more stringent junctional binding affinity limit of 1500 nM, with a maximum clip of 3 aa, representing a more challenging design scenario. Finally, the IC1500_Clip5 setting used the same 1500 nM junctional binding affinity limit, but allowed clipping of up to 5 aa, representing a scenario which is more permissive than IC1500 setting.

Across all three settings, pVACvector version 5 achieved shorter runtimes and higher design success rates than version 4 (**Sup Table 2**, **Figure 7**). The runtime improvements were substantial: under the IC1500 setting, version 5 reduced average runtime from 334.6 to 131.2 minutes - a 61% reduction, and median runtime from 201.9 to 97.2 minutes. Under the IC1500_Clip5 setting, average runtime decreased from 326 to 134.1 minutes (59% reduction). Even under the default setting, where both versions performed relatively quickly, version 5 modestly reduced average runtime from 70.9 to 57.5 minutes (**Sup Table 2**).

Beyond runtime, version 5 also produced more biologically optimized vector designs. In one test case (G104, IC1500_Clip5 setting), version 4 generated a solution that inadvertently clipped amino acids within the predicted MHC binding core of one peptide, compromising the integrity of a key neoantigen sequence. Version 5, by contrast, identified a solution that fully preserved the binding core, demonstrating that the improved algorithm not only finds solutions faster but finds better ones (**Figure 7, Supplementary Figure 6**). Taken together, these results support pVACvector version 5 as a substantially more efficient and reliable tool for personalized DNA vaccine vector design.

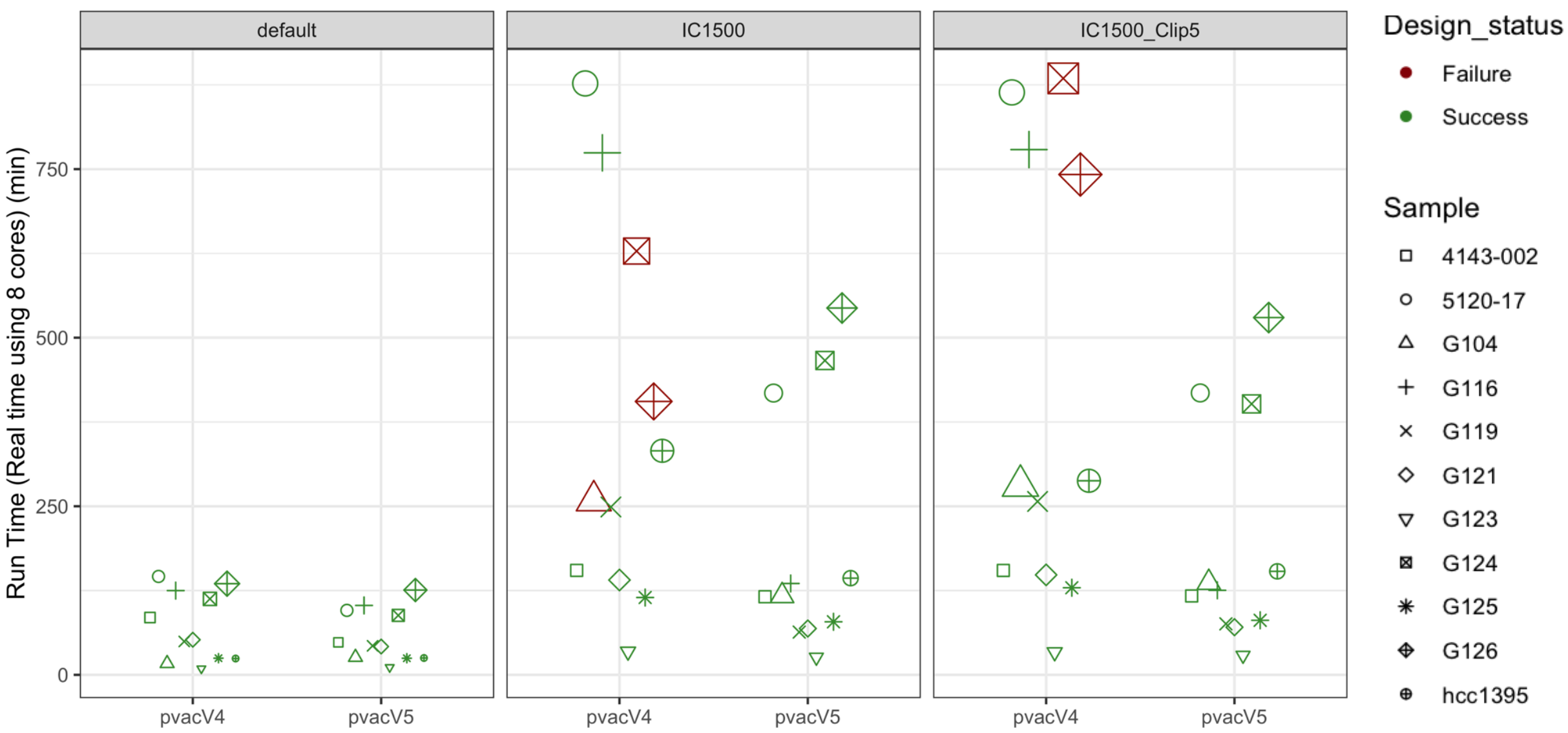

**Figure 7. Benchmarking pVACvector version 4 and version 5 algorithms demonstrates shorter runtime and more successful designs in version 5.** Runtime and design status for pVACvector version 4 (pvacV4) and version 5 (pvacV5) across 11 cases and 3 different settings. Default setting has junctional binding affinity (IC50 score) limit of 500 nM, and max clipped peptide of 3 amino acids (aa). IC1500 setting has junctional binding affinity limit of 1500 nM and max clipped peptide of 3 aa. IC1500_Clip5 has junctional binding affinity limit of 1500 nM and max clipped peptide of 5 aa. Each individual point shows runtime (in minutes) of each run. A case is colored green if pVACvector successfully produced a linear multi-epitope design with no junctional epitopes and colored red if pVACvector failed to come up with a design.

## Demonstration of pVACtools on non-human species

To demonstrate the utility of pVACtools for analysis of non-human data, we applied pVACtools to predict neoantigens from exome and RNA-seq data of a tumor sarcoma mouse model (T3[61]) against 129S6 background mice (**Figure 8A**). Our pipeline identified 10,717 simple somatic variants. Of these, pVACseq identified 3,308 variants as SNV/indels in coding regions, estimated 1,561 as well-expressed (Allelic expression > 0.25), predicted 879 as good MHC binders, and labeled 332 variants with tier ‘Pass’. Within the above mentioned 10,717 variants, RegTools identified 1,215 as putative cis-splicing variants spanning ±10 nucleotides around the intron-exon boundary, and of these pVACsplice further identified 212 variants as potential candidates (with novel junctions robustly supported at >10 reads RNA level, and passing other pVACsplice pre-filters), 201 as well-expressed (as above), 84 as potential good binders, 59 tiered ‘Pass’ (**Figure 8B**).

To validate pVACtools predicted neoantigens, we performed mass spectrometry-based identification of MHC-bound peptides (immunopeptidomics) on the murine tumor sample, using two data acquisition approaches: DDA and DIA (**Figure 8A**). We used pVACseq and pVACsplice `generate protein fasta` function to create mutated peptide sequences, then combined these sequences with reference proteome, contaminants and decoy sequences to create a database for mass spec search in FragPipe[62,63]. The search revealed thousands of unique peptides with the expected MHC-binding motif identified by unsupervised clustering with MHCMotifDecon[64] (**Supplementary Figure 7**). The search also confirmed neoantigens arising from 4 SNVs (**Figure 8C**). Notably, all mass spectrometry validated peptide sequences corresponded to the Best Peptide nominated by pVACtools for their respective mutations. These peptides all have good binding affinity prediction (IC50 < 500nM, IC50 %ile < 1%), and their associated mutations are well supported (high DNA VAF, RNA VAF, RNA and allele expression) (**Supplementary Figure 8**). These candidates are all assigned within the “Pass” Tier in the pVACseq aggregate report and pVACview. Of note, the Lama4 G1254V peptide, ranked #1 by pVACseq, is a previously well-characterized, immunodominant antigen that has been shown to induce antitumor immunity against T3 tumors when administered as a synthetic long peptide vaccine[65,66]. The other three candidates (Vkorc1, Mphosph8, Zc3h13) are novel and have never been characterized. Taken together, these results show that pVACtools in silico neoantigen prediction is capable of predicting bona fide MHC-presented peptides, and pVACtools prioritization scheme can reliably assist selection of epitopes for downstream application in tumor models.

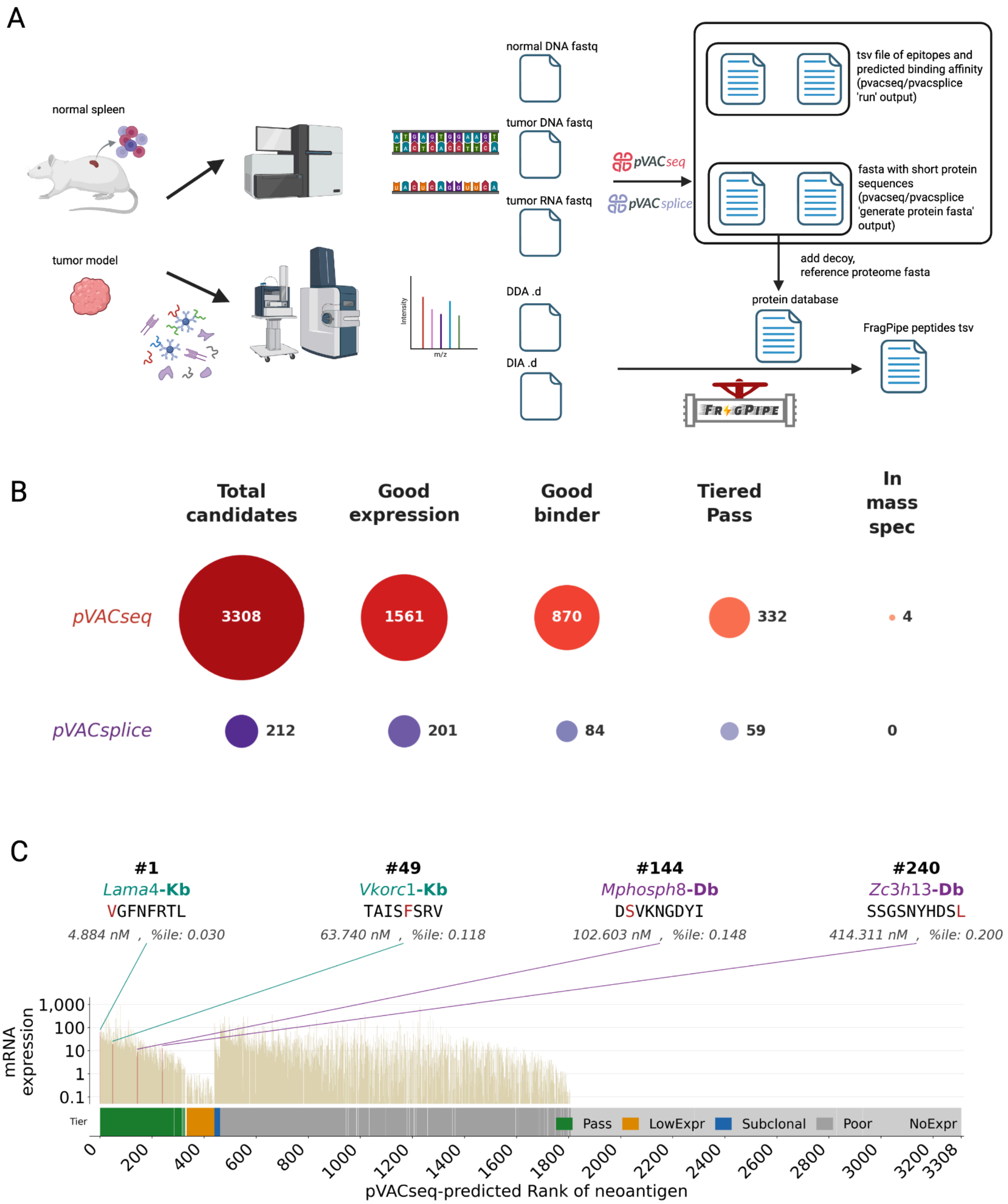

**Figure 8. pVACtools suite proposes putative neoantigens that are validated in immunopeptidomics data for T3 murine sarcoma model.** (A) Experimental schematic. Tumor from T3 mice and normal spleen cells from 129S6 mice were sent for WES, RNAseq and mass spectrometry. The raw sequencing data were analysed with a workflow to predict neoantigens using pVACseq and pVACsplice. The pvacseq/pvacsplice 'run' commands provided neoantigen binding predictions, as well as nominated the best overall neoantigen candidate. The pvacseq/pvacsplice 'generate protein fasta' commands were used to generate a list of peptides with desired flanking amino acid length, which were

used as the protein database for spectral matching via FragPipe. (B) Summary of neoantigen candidates counts. (C) Ranked pVACseq in-silico predicted neoantigens. Bottom: Tier of predicted neoantigen candidates (Pass, Low expression, Suclonal, Poor, Not Expressed); Middle: RNA expression (in TPM) of variant associated with the candidates; Top: Epitopes validated by immunopeptidomics are highlighted. The mutated amino acids for MS-validated epitopes are colored in red. Predicted binders for H-2-Kb allele are colored in teal, predicted binder for H-2-Db alleles are colored in purple.

# Materials and methods

## Implementation

pVACtools is written largely in Python3 with pVACview written in R Shiny. The individual tools are implemented as separate command line entry points that can be run using the `pvacseq`, `pvacfuse`, `pvacsplice`, `pvacvector`, and `pvacview` commands for each respective tool. Each of these command line entry points contains a `run` subcommand to run the main pipelines described in this paper. Additionally, many of the individual steps in the main pipelines are also available as subcommands to allow for users to, e.g. filter their result files with different thresholds, regenerate the aggregated report with different tiering parameters, or manually execute optional steps that may have been omitted from a main pipeline run like reference proteome matching, cleavage site prediction, or stability predictions. Additional downstream tools are available as subcommands under each command line entry point. These include functionality for generating protein fasta files with a specific desired flanking sequence length, generating a peptide ordering spreadsheet, etc.. Helper commands for listing supported alleles and prediction algorithms are also available under each command line entry point. In addition to the already mentioned command line entry points, a generic `pvactools` entry point includes tools for downloading CWL and WDL workflow files as well as the `pvactools compare` command to run the pVACcompare tool which identifies differences between output files resulting from different pVACtools versions or different parameters specified during a user's run.

The code test suite is implemented using the Python unittest framework and GitHub integration tests are run using GitHub actions. Code changes are integrated using GitHub pull requests (https://github.com/griffithlab/pVACtools/pulls). Feature additions, user requests, and bug reports are managed using the GitHub issue tracking (https://github.com/griffithlab/pVACtools/issues). User documentation is written using the reStructuredText (RST) markup language and the Sphinx documentation framework (sphinx-doc.org). Documentation is hosted on Read the Docs (readthedocs.org) and can be viewed at pvactools.org. Additionally, the OTTR framework is used to create educational courses, such as course.pvactools.org and workflow-course.pvactools.org.

## Running RegTools and pVACsplice on tumor cohorts

GBM/BrMet cohort:

Sample and data generation details for the GBM/BrMEt cohort and sequencing data were described in a previous publication[59]. Cis-splicing events were detected with RegTools version 0.5.2, using default settings, with command `regtools cis-splice-effects identify`. Splice junctions absent from normal tissues in GTEx[60] version 10 were identified using custom scripts. GTEx junctions are downloaded from GTEx website (at https://storage.googleapis.com/adult-gtex/bulk-gex/v10/rna-seq/GTEx_Analysis_v10_STARv2.7.10a_junctions.gct.gz). Neoantigens for respective patient HLA types were predicted with pVACsplice ver 5.3.0 (from DockerHub image 'griffithlab/pvactools:5.3.0') with command `pvacsplice run` using the following options: `-g -k -t <thread number> --iedb-install-directory /opt/iedb --normal-sample-name <normal sample name>`.

SCLC cohort:
Similar to the GBM/BrMet cohort, SCLC cohort cis-splicing events were predicted with Regtools v.1.0.0 and pVACsplice v.5.0.0.

## Benchmark pVACvector old and new algorithms

A new pVACvector graph building algorithm (v.5.2.0) was benchmarked against the old algorithm (v.4.4.1) using the corresponding versioned 'griffithlab/pvactools' docker containers from DockerHub.
pVACvector was run with the `pvacvector run` command using the following options: `--binding-threshold $BINDING_THRESHOLD -k -t 8 --iedb-install-directory /opt/iedb --max-clip-length $MAXCLIPLENGTH`. The $BINDING_THRESHOLD is 500 (nM) for Default mode and 1500 (nM) for IC1500 and IC1500_Clip5 modes. The $MAXCLIPLENGTH is not specified (and thereby set at default value of 3) for Default and IC1500 modes, and set to 5 for IC1500_Clip5 mode. Benchmarking was performed with 8 prediction algorithms: NetMHCpan,NetMHC,NetMHCcons,PickPocket,SMM,SMMPMBEC,MHCflurry,MHCnuggetsI. Patient's HLAs (from either Optitype or HLAtyping clinical report) were used for the analysis. Runtime was captured using the `time` command in unix. For runtime comparison, real time in minutes was used. The comparison plot was created with ggplot2 v3.5.1.

## Sequencing data acquisition, analysis, and neoantigen prediction on mouse model data

Tumors from T3 mice (MCA-induced sarcomas, from frozen stock[67]) and normal spleen cells from 129S6 mice (to match the sex and strain of T3 tumors) were processed then sent for WES, RNAseq (as previously described[61,65]) and mass spectrometry-based immunopeptidomics as described below.

WES libraries were prepared using the Whole Exome-KAPA Hyper amplified kit and sequenced on a NovaSeqX platform (NovaSeqXPlus flowcell, 300 cycles). RNA libraries were prepared using the Bulk RNAseq-Watchmaker RNA with depletion kit and sequenced on a NovaSeqX platform (NovaSeqXPlus flowcell, 300 cycles).

The raw sequencing data were processed through a workflow optimized for non-human data, which performs sequencing read alignment to genome, somatic variant calling, RNA-seq analysis, and finally predicts neoantigen (for MHC alleles H-2-Kb, H-2-Db) with pVACseq and pVACsplice. The workflow WDLs for alignment, variant calling, RNA abundance estimation

steps are available at
https://github.com/wustl-oncology/analysis-wdls/blob/main/definitions/nonhuman_immuno.wdl. RegTools (pre-requisite step for pVACsplice) was run with intronic variants within 10 bases and exonic variants within 10 bases of the exon edge (setting `-i 10 -e 10`). The pvacseq/pvacsplice `run` command provides neoantigen binding prediction, as well as nominates the best neoantigen candidate.

## Immunopeptidomics data acquisition and analysis for mouse model data

### Isolation of MHC-I peptidomes

For each of the independent analyses, the frozen cell pellets were thawed on the isolation day and suspended in fresh lysis buffer (1.2% CHAPS (MilliporeSigma, Cat #: 220201), 10 mM iodoacetamide (MilliporeSigma, Cat #: I6125), and Pierce Protease Inhibitor mini tablets (ThermoFisher, Cat #: A32955) in PBS). The suspension was rocked for 1 h at 4 °C. The cell lysate was then centrifuged at 20,000×*g* for 30 min at 4 °C. To eliminate non-specific binding of peptides, the supernatant was first incubated with polyclonal mouse IgG (Leinco, Cat #: N229; 300 ug antibody per sample) bound to Sepharose 4B (MilliporeSigma, Cat #: C9142) at 4 °C for 30 min. The unbound material containing peptide-MHC-I complexes was collected and added to a tube containing PBS-washed sepharose conjugated to the anti-mouse H-2K(b) antibody (Y-3, Leinco Cat#: Y200; 300 ug per sample). This mixture was incubated at 4 °C overnight. The sepharose was applied to a column and washed four times as follows: 10 ml 150 mM NaCl and 20 mM Tris (pH 7.4), 10 ml 400 mM NaCl and 20 mM Tris (pH 7.4), 10 ml 150 mM NaCl and 20 mM Tris (pH 7.4), and 10 ml 20 mM Tris (pH 8.0). Peptides were eluted with 0.2% trifluoroacetic acid (MilliporeSigma, Cat #: T6508) and dried using a SpeedVac. The unbound material from the H-2K(b) peptide isolations, containing other peptide-MHC-I complexes, was collected and added to a tube containing PBS-washed sepharose conjugated to the anti-mouse H-2D(b) antibody (B22/249, Leinco Cat#: B341; 300 ug antibody per sample). This mixture was incubated at 4 °C overnight. The sepharose was applied to a column and washed four times as follows: 10 ml 150 mM NaCl and 20 mM Tris (pH 7.4), 10 ml 400 mM NaCl and 20 mM Tris (pH 7.4), 10 ml 150 mM NaCl and 20 mM Tris (pH 7.4), and 10 ml 20 mM Tris (pH 8.0). Peptides were eluted with 0.2% trifluoroacetic acid and dried using a SpeedVac. Eluted peptides were resuspended and passed over detergent removal spin columns (Pierce, Cat #: 87777) to remove traces of remaining detergent, and then cleaned using C18 Spin Columns (Pierce, Cat #: 89870). Two C18 spin column cleanup steps were performed for each sample: peptides were eluted from the first C18 column with 75% acetonitrile (Supelco, Cat#: 1.00029.1000)/0.1% formic acid (Pierce, Cat#: 28905), reapplied to a second C18 spin column, and eluted from the second C18 column with 60% acetonitrile. Peptides were then dried down for mass spectrometry analysis.

### Ultra high-performance liquid chromatography mass spectrometry – timsTOF Ultra 2

The peptides were analyzed using trapped ion mobility time-of-flight mass spectrometry (PMID30385480). Peptides were separated using a nano-ELUTE 2 liquid chromatograph (Bruker, Bremen, Germany) interfaced to a timsTOF Ultra 2 mass spectrometer (Bruker) with a modified nano-electrospray source (CaptiveSpray 2, Bruker). The samples were reconstituted in 5 µl of 0.1% (vol/vol) formic acid (FA) in water, and 4 µl were injected onto a 75 µm i.d. × 25 cm

PepSep Ultra column (Bruker) connected to a 10 µm emitter (Bruker). The column temperature was set to 50 °C. The column was equilibrated using constant pressure (800 bar) with 5 column volumes of solvent A (0.1% (vol/vol) FA). Sample loading was performed at constant pressure (800 bar) at a volume of 2 sample pick-up volume plus 2 µl. The peptides were eluted using one column separation mode with a flow rate of 250 nl/min and using solvents A (0.1% (vol/vol) FA) and B (0.1% (vol/vol) FA/Acetonitrile): solvent A containing 2% B increased to 20% B over 40 min, to 37% B over 5 min, to 95% B over 1 min, and constant 95% B for 4 min. The mass spectrometer was operated in the high sensitivity DDA PASEF mode (PMID30385480). Ion mobility range was set to 0.64 – 1.65 Vs/cm2 (0.64-1.64 Vs/cm2 for MHC-II) with ramp time 100 ms and accumulation time 100 ms. MS1 and MS2 spectra were recorded from m/z 100 to 1700. The number of PASEF ramps was set to 5, and the intensity threshold was set to 1,000, and the target intensity was 12,000. Deflection 0 Delta was set at 30.0 V (70.0 V for MHC-II). Fast quadrupole switching was used (measuring time 2.00ms, switching time 1.20ms). Quadrupole isolation was set to 2 Th at <700 m/z and 3 Th at 800 m/z. The samples were reconstituted in 5 µl of 0.1% (vol/vol) formic acid (FA) in water, and 4 µl were injected onto a 75 µm i.d. × 25 cm PepSep Ultra column (Bruker) connected to a 10 µm emitter (Bruker). The column temperature was set to 50 °C. The column was equilibrated using constant pressure (800 bar) with 5 column volumes of solvent A (0.1% (vol/vol) FA). Sample loading was performed at constant pressure (800 bar) at a volume of 2 sample pick-up volume plus 2 µl. The peptides were eluted using one column separation mode with a flow rate of 250 nl/min and using solvents A (0.1% (vol/vol) FA) and B (0.1% (vol/vol) FA/Acetonitrile): solvent A containing 2% B increased to 20% B over 40 min, to 37% B over 5 min, to 95% B over 1 min, and constant 95% B for 4 min. The MS1 and MS2 spectra were recorded from m/z 100 to 1700. The collision energy was ramped stepwise as a function of increasing ion mobility: 20 eV at 0.60 Vs/cm2 and 59 eV 1.60 Vs/cm2, and MS2 spectra were acquired with moderate denoising. The TIMS elution voltage was calibrated linearly using the Agilent ESI-L Tuning Mix (m/z 622, 922,1222).

**Data analysis - FragPipe**

Identification of pVAC predicted peptides in mass spectrometry data was performed in FragPipe version 23.1. pVACseq and pVACsplice `generate protein fasta` output fastas were combined with mouse reference proteome (Uniprot ID UP000000589), as well as decoy sequences and contaminants (generated by FragPipe) to create a single fasta file used as input database for the mass spectrometry search. The database search was performed using an altered version of FragPipe ‘nonspecific HLA diaPASEF’ workflow, with +/- 20 ppm for parent tolerance, 30 ppm for fragment ion tolerance, C (cysteine) 57.02146 and no ambiguous amino acids (B, J, O, U, X, Z) in fixed modifications.

# Discussion

The growing clinical momentum behind personalized neoantigen vaccines, demonstrated by recent randomized trial results and the expansion of multi-epitope vaccine platforms[6–9] has underscored the need for robust, comprehensive computational tools to support the full neoantigen discovery and therapy production workflow. Here, we describe updates to the pVACtools suite across three major areas: expansion of the targetable neoantigen repertoire through the introduction of pVACsplice, enhancement of neoantigen quality assessment features, and improvements to neoantigen selection and vaccine design. Additionally, we have expanded support for exploratory neoantigen sources (pVACbind) and non-human samples, updated the neoantigen visualization tool pVACview, and added runs comparison tool pVACcompare. Taken together, these updates reflect an evolving understanding of what constitutes a high-quality neoantigen candidate and our efforts to embed that understanding directly into the computational workflow.

The first major advancement is the expansion of pVACtools' covered mutational repertoire. While the previous version of pVACtools supported neoantigen prediction from SNVs, indels, and gene fusions, a substantial fraction of tumor-specific neoantigens arise from aberrant splicing events - a source that has historically been underserved by existing pipelines. pVACsplice addresses this by predicting neoantigens from tumor-specific alternative splicing junctions identified from paired whole-exome and RNA sequencing data. Several other tools have explored splicing neoantigen prediction. Tools such as ASNEO[68], IRIS[69], ISOTOPE[70], NeoSplice[71], SNAF-T[72] are distinct from pVACsplice as they predict splicing neoantigens by relying solely on tumor RNAseq data and attempt to filter for tumor-specific alternative splicing events either by incorporation of matched normal RNAseq, or natively filtering out junctions in normal tissue reference datasets such as GTEx. While these filtering strategies can reduce noise, confidently establishing that a splicing event is truly tumor-specific remains challenging. Even with a matched normal RNA comparator, obtaining a perfectly representative reference is difficult in practice: the normal sample must match not only the tissue type of the tumor, but ideally its transcriptional cell states - a bar that is rarely met given tumor heterogeneity and other context-dependent shifts in cell identity. Corpus-based references such as GTEx introduce their own limitations, including batch effects and incomplete representation of the relevant normal cell states. In contrast, pVACsplice takes a deliberately conservative approach by anchoring splicing events to somatic DNA variants, providing an orthogonal line of evidence that the event is genuinely tumor-derived rather than a low-level splice isoform present in normal tissue. Splice2neo[73] generates the mutated peptide sequences from novel junctions, but doesn't directly predict neoantigen binding affinity to MHC like pVACsplice, and requires a downstream tool to perform prediction. SpliceMutr[74] predicts splicing neoantigens at the cohort-level, and unlike pVACsplice, isn't suitable for n-of-1 cases. SPLICE-neo[75], similar to pVACsplice, also explores cis-splicing neoantigens, however, the tool does not support splicing neoantigens by non-canonical splice sites, as opposed to RegTools-pVACsplice which supports all cis-splicing mutations. Moreover, pVACsplice is also distinguished from all aforementioned tools by its integration of multiple MHC binding, presentation, and immunogenicity prediction algorithms within a unified framework, its incorporation of DNA and RNA variant allele frequency and expression data for candidate prioritization, and its seamless connection to downstream vaccine manufacturing support, and neoantigen exploration using a well-established sequencing-to-neoantigen-prediction workflow (via ImmunoNX[35]). Notably, because matched normal RNA sequencing from the tumor's tissue of origin is rarely available in clinical settings, pVACsplice's variant-anchored approach - which prioritizes novel junctions supported by nearby somatic variants - provides a practical strategy to enrich for tumor-specific events without requiring normal RNA data.

The second major advancement is the expansion of neoantigen quality assessment features. The previous version of pVACtools integrated a comprehensive set of MHC binding affinity prediction algorithms for both MHC class I and class II, along with peptide processing (NetChop) and peptide-MHC stability (NetMHCstabpan). In version 5, we have added support for presentation scoring (MHCflurryEL, NetMHCpanEL, BigMHC-EL for class I; NetMHCIIpanEL for class II) and immunogenicity prediction (BigMHC_IM, DeepImmuno), providing a more complete picture of the likelihood that a candidate peptide will be presented and recognized by T cells. While experimental data used for training immunogenicity predictors is still limited in size, and caution in interpretation is warranted, immunogenicity scores are nonetheless useful in many cases. We have also added several features to assess neoantigen dissimilarity to self: reference proteome similarity search via BLASTp and anchor residue prediction to identify cases where the mutation falls at an MHC contact position rather than a TCR-facing position. These features collectively enable users to move beyond binding affinity alone and apply a more biologically grounded prioritization strategy.

The third advancement spans neoantigen selection and vaccine manufacturing support. For selection, pVACtools now incorporates an internal ranking scheme across pVACseq, pVACfuse, and pVACsplice outputs, enabling rapid review of large candidate lists. To enable neoantigen feature exploration, pVACview provides an interactive visualization environment that supports both pVACtools outputs and those from third-party pipelines. We note that upstream tools - including variant callers and expression quantification methods - are imperfect and advise users to treat pVACtools rankings as a prioritization guide rather than a definitive filter, and to visually inspect variant calls using tools such as IGV as part of their quality control process. For vaccine manufacturing, the pVACvector algorithm has been substantially improved in version 5. Benchmarking across 11 vaccine real world design cases demonstrated that the new algorithm achieves a higher design success rate and shorter runtimes compared to version 4 across all tested parameter settings. pVACvector version 5 also produces more biologically optimized vector sequences - for example, avoiding inadvertent clipping of MHC binding cores that was observed with the version 4 algorithm. In addition to improvements in vector design, pVACtools also offers additional assistance for peptide vaccine design, through a new peptide ordering form feature in pVACseq, as well as various criteria for peptide synthesis. These improvements allow pVACtools to better support clinical vaccine design workflows.

Looking ahead, several directions are under consideration for future development of pVACtools. On the neoantigen source front, neoantigens arising from structural variants, RNA editing events, transposable element expression, viral integration, and post-translational modifications such as phosphorylation and glycosylation represent areas of active interest for expanding the tool's coverage. Beyond sequencing-based approaches, the integration of proteomics and immunopeptidomics data represents a valuable complementary strategy: tools such as ProGeo-neo[76], NeoDisc[77], NeoFlow[78], and pyGeno[79] have demonstrated the feasibility of using customized protein databases derived from tumor sequencing data to search for mutant peptides in mass spectrometry data. Such integration can provide experimental evidence that predicted peptides are processed and presented in vivo. We note, however, that the absence of a predicted peptide in mass spectrometry data should not be interpreted as evidence against its candidacy: the sensitivity of current immunopeptidomics workflows remains limited[80]. We look forward to incorporating proteomics data integration into future versions of pVACtools as these technologies mature. Also on the neoantigen source front, in pVACsplice, we plan to add an additional module that supports generic splicing neoantigen prediction (not limited to cis-splicing neoantigens). One aspect of neoantigen quality assessment that pVACtools does not yet fully support is tumor variant clonality estimation. pVACtools uses variant allele frequency (VAF) and tumor purity as proxies for clonality. Support for cancer cell fraction (CCF) will be added to account for local copy number status and provide a better estimate of the proportion of tumor cells harboring a given variant. Support for additional prediction algorithms (ImmuScope, MixMHCpred, PRIME, MixMHC2pred) is also planned. To facilitate a more fair comparison of prediction scores across algorithms, we will provide internally calculated, normalized per-algorithm percentile scores in the future. These normalized percentile scores are generated by ranking the peptide of interest against a harmonized/common set of training data, allowing a fair ranking for the peptide of interest. Future versions will also provide matched wildtype peptides in pVACfuse and pVACsplice to allow dissimilarity to self assessment in these modules. This change will also enable support for pVACfuse and pVACsplice outputs in pVACview to support more comprehensive neoantigen review and selection in these pipelines. Upcoming pVACtools version will also take into account presentation and immunogenicity score in neoantigen tiering. Finally, we plan to release a machine-learning-based ranking model to complement the current rule-based neoantigen prioritization scheme.

# Tool availability and resources

The pVACtools source code is available at https://github.com/griffithlab/pVACtools with versioned releases made available on the Python package index (PyPi), DockerHub, and GitHub, and Zenodo. Note that major version releases are not backward compatible; users are advised to use the latest version to access the suite's full functionality.

pVACtools github: https://github.com/griffithlab/pVACtools
pVACview online server: https://www.pvacview.org/
pVACcompare github: https://github.com/griffithlab/pVACcompare
pVACtools documentation: https://pvactools.readthedocs.io/en/latest/index.html
pVACtools Zenodo DOI: https://doi.org/10.5281/zenodo.3355660
pVACtools docker: https://hub.docker.com/r/griffithlab/pvactools
pVACtools Pypi: https://pypi.org/project/pvactools/
Wdl workflow: https://github.com/wustl-oncology/analysis-wdls
ImmunoNX protocol github: https://github.com/griffithlab/ImmunoNX_protocol
pVACtools course: https://course.pvactools.org/introduction.html
pVACtools youtube: https://www.youtube.com/@pvactools8083

# Supplementary Figures

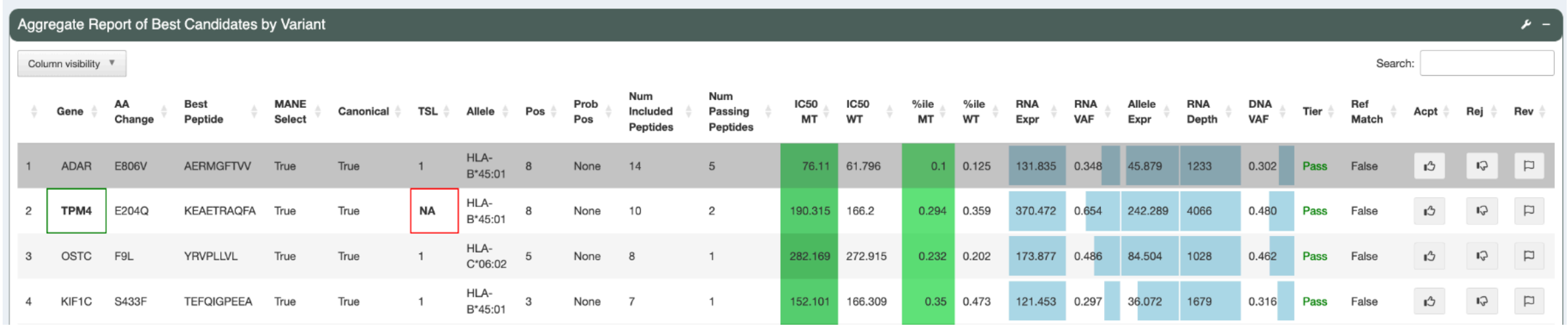

Aggregate Report of Best Candidates by Variant

|  | Gene | AA Change | Best Peptide | MANE Select | Canonical | TSL | Allele | Pos | Prob Pos | Num Included Peptides | Num Passing Peptides | IC50 MT | IC50 WT | %ile MT | %ile WT | RNA Expr | RNA VAF | Allele Expr | RNA Depth | DNA VAF | Tier | Ref Match | Acpt | Rej | Rev |
|---|---|---|---|---|---|---|---|---|---|---|---|---|---|---|---|---|---|---|---|---|---|---|---|---|---|
| 1 | ADAR | E806V | AERMGFTVV | True | True | 1 | HLA-B*45:01 | 8 | None | 14 | 5 | 76.11 | 61.796 | 0.1 | 0.125 | 131.835 | 0.348 | 45.879 | 1233 | 0.302 | Pass | False |  |  |  |
| 2 | TPM4 | E204Q | KEAETRAQFA | True | True | NA | HLA-B*45:01 | 8 | None | 10 | 2 | 190.315 | 166.2 | 0.294 | 0.359 | 370.472 | 0.654 | 242.289 | 4066 | 0.480 | Pass | False |  |  |  |
| 3 | OSTC | F9L | YRVPLLVL | True | True | 1 | HLA-C*06:02 | 5 | None | 8 | 1 | 282.169 | 272.915 | 0.232 | 0.202 | 173.877 | 0.486 | 84.504 | 1028 | 0.462 | Pass | False |  |  |  |
| 4 | KIF1C | S433F | TEFQIGPEEA | True | True | 1 | HLA-B*45:01 | 3 | None | 7 | 1 | 152.101 | 166.309 | 0.35 | 0.473 | 121.453 | 0.297 | 36.072 | 1679 | 0.316 | Pass | False |  |  |  |

**Supplementary Figure 1. The pVACview main module interface, which summarizes data from the pVACseq aggregate report and other pVACseq outputs in an interactive manner**. Each row shows one variant, with the proposed representative candidate (Best Peptide), along with its Tier. Selecting each row allows users to explore extensive further details on the top prioritized peptide for each somatic variant, additional ranked peptides, transcript annotation information, individual algorithm predictions, etc.

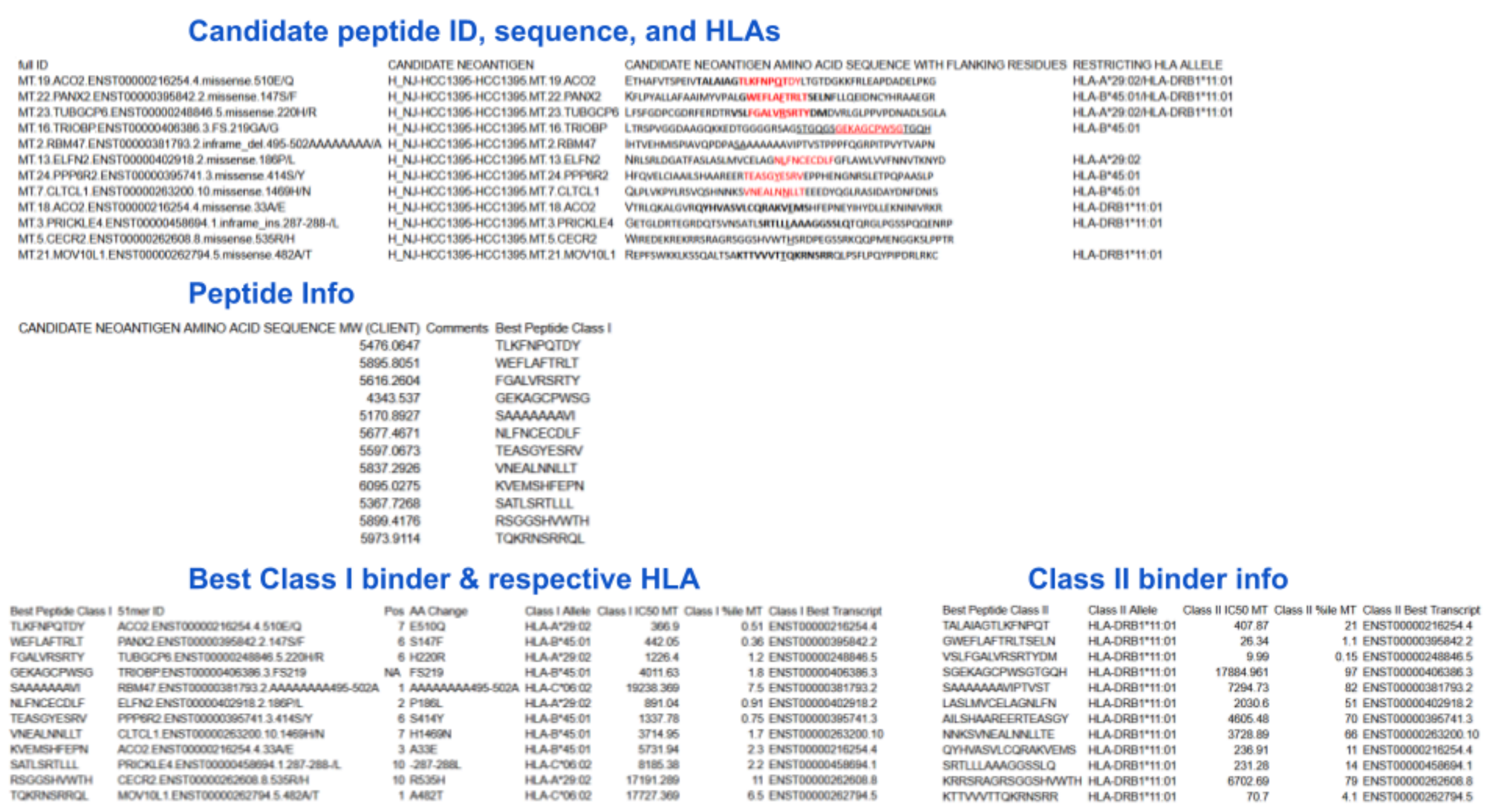

Candidate peptide ID, sequence, and HLAs

| full ID | CANDIDATE NEOANTIGEN | CANDIDATE NEOANTIGEN AMINO ACID SEQUENCE WITH FLANKING RESIDUES | RESTRICTING HLA ALLELE |
|---|---|---|---|
| MT.19.ACO2.ENST00000216254.4.missense.510E/Q | H_NJ-HCC1395-HCC1395.MT.19.ACO2 | ETHAFVTSPEIVTALAIAGTLKFNPQTDYLTGTDGKKFRLEAPDADELPKG | HLA-A*29:02/HLA-DRB1*11:01 |
| MT.22.PANX2.ENST00000395842.2.missense.147S/F | H_NJ-HCC1395-HCC1395.MT.22.PANX2 | KFLPYALLAFAAIMYVPALGWEFLAFTRLTSELNFLLQEIDNCYHRAAEGR | HLA-B*45:01/HLA-DRB1*11:01 |
| MT.23.TUBGCP6.ENST00000248846.5.missense.220H/R | H_NJ-HCC1395-HCC1395.MT.23.TUBGCP6 | LFSFGDPCGDRFERDTRVSLFGALVRSRTYDMDVRLGLPPVPDNADLSGLA | HLA-A*29:02/HLA-DRB1*11:01 |
| MT.16.TRIOBP.ENST00000406386.3.FS.219GA/G | H_NJ-HCC1395-HCC1395.MT.16.TRIOBP | LTRSPVGGDAAGQKKEDTGGGGRSAGSTGQGSGEKAGCPWSGTGQH | HLA-B*45:01 |
| MT.2.RBM47.ENST00000381793.2.inframe_del.495-502AAAAAAAA/A | H_NJ-HCC1395-HCC1395.MT.2.RBM47 | IHTVEHMISPIAVQPDPASAAAAAAAVIPTVSTPPPFQGRPITPVYTVAPN |  |
| MT.13.ELFN2.ENST00000402918.2.missense.186P/L | H_NJ-HCC1395-HCC1395.MT.13.ELFN2 | NRLSRLDGATFASLASLMVCELAGNLFNCECDLFGFLAWLVVFNNVTKNYD | HLA-A*29:02 |
| MT.24.PPP6R2.ENST00000395741.3.missense.414S/Y | H_NJ-HCC1395-HCC1395.MT.24.PPP6R2 | HFQVELCIAAILSHAAREERTEASGYESRVEPPHENGNRSLETPQPAASLP | HLA-B*45:01 |
| MT.7.CLTCL1.ENST00000263200.10.missense.1469H/N | H_NJ-HCC1395-HCC1395.MT.7.CLTCL1 | QLPLVKPYLRSVQSHNNKSVNEALNNLLTEEEDYQGLRASIDAYDNFDNIS | HLA-B*45:01 |
| MT.18.ACO2.ENST00000216254.4.missense.33A/E | H_NJ-HCC1395-HCC1395.MT.18.ACO2 | VTRLQKALGVRQYHVASVLCQRAKVEMSHFEPNEYIHYDLLEKNINIVRKR | HLA-DRB1*11:01 |
| MT.3.PRICKLE4.ENST00000458694.1.inframe_ins.287-288-/L | H_NJ-HCC1395-HCC1395.MT.3.PRICKLE4 | GETGLDRTEGRDQTSVNSATLSRTLLLAAAGGSSLQTQRGLPGSSPQQENRP | HLA-DRB1*11:01 |
| MT.5.CECR2.ENST00000262608.8.missense.535R/H | H_NJ-HCC1395-HCC1395.MT.5.CECR2 | WIREDEKREKRRSRAGRSGGSHVWTHSRDPEGSSRKQQPMENGGKSLPPTR |  |
| MT.21.MOV10L1.ENST00000262794.5.missense.482A/T | H_NJ-HCC1395-HCC1395.MT.21.MOV10L1 | REPFSWKKLKSSQALTSAKTTVVVTTQKRNSRRQLPSFLPQYPIPDRLRKC | HLA-DRB1*11:01 |

Peptide Info

| CANDIDATE NEOANTIGEN AMINO ACID SEQUENCE MW (CLIENT) | Comments | Best Peptide Class I |
|---|---|---|
| 5476.0647 |  | TLKFNPQTDY |
| 5895.8051 |  | WEFLAFTRLT |
| 5616.2604 |  | FGALVRSRTY |
| 4343.537 |  | GEKAGCPWSG |
| 5170.8927 |  | SAAAAAAAVI |
| 5677.4671 |  | NLFNCECDLF |
| 5597.0673 |  | TEASGYESRV |
| 5837.2926 |  | VNEALNNLLT |
| 6095.0275 |  | KVEMSHFEPN |
| 5367.7268 |  | SATLSRTLLL |
| 5899.4176 |  | RSGGSHVWTH |
| 5973.9114 |  | TQKRNSRRQL |

Best Class I binder & respective HLA

| Best Peptide Class I | 51mer ID | Pos | AA Change | Class I Allele | Class I IC50 MT | Class I %ile MT | Class I Best Transcript |
|---|---|---|---|---|---|---|---|
| TLKFNPQTDY | ACO2.ENST00000216254.4.510E/Q | 7 | E510Q | HLA-A*29:02 | 366.9 | 0.51 | ENST00000216254.4 |
| WEFLAFTRLT | PANX2.ENST00000395842.2.147S/F | 6 | S147F | HLA-B*45:01 | 442.05 | 0.36 | ENST00000395842.2 |
| FGALVRSRTY | TUBGCP6.ENST00000248846.5.220H/R | 6 | H220R | HLA-A*29:02 | 1226.4 | 1.2 | ENST00000248846.5 |
| GEKAGCPWSG | TRIOBP.ENST00000406386.3.FS219 | NA | FS219 | HLA-B*45:01 | 4011.63 | 1.8 | ENST00000406386.3 |
| SAAAAAAAVI | RBM47.ENST00000381793.2.AAAAAAAA495-502A | 1 | AAAAAAAA495-502A | HLA-C*06:02 | 19238.369 | 7.5 | ENST00000381793.2 |
| NLFNCECDLF | ELFN2.ENST00000402918.2.186P/L | 2 | P186L | HLA-A*29:02 | 891.04 | 0.91 | ENST00000402918.2 |
| TEASGYESRV | PPP6R2.ENST00000395741.3.414S/Y | 6 | S414Y | HLA-B*45:01 | 1337.78 | 0.75 | ENST00000395741.3 |
| VNEALNNLLT | CLTCL1.ENST00000263200.10.1469H/N | 7 | H1469N | HLA-B*45:01 | 3714.95 | 1.7 | ENST00000263200.10 |
| KVEMSHFEPN | ACO2.ENST00000216254.4.33A/E | 3 | A33E | HLA-B*45:01 | 5731.94 | 2.3 | ENST00000216254.4 |
| SATLSRTLLL | PRICKLE4.ENST00000458694.1.287-288-/L | 10 | -287-288L | HLA-C*06:02 | 8185.38 | 2.2 | ENST00000458694.1 |
| RSGGSHVWTH | CECR2.ENST00000262608.8.535R/H | 10 | R535H | HLA-A*29:02 | 17191.289 | 11 | ENST00000262608.8 |
| TQKRNSRRQL | MOV10L1.ENST00000262794.5.482A/T | 1 | A482T | HLA-C*06:02 | 17727.369 | 6.5 | ENST00000262794.5 |

Class II binder info

| Best Peptide Class II | Class II Allele | Class II IC50 MT | Class II %ile MT | Class II Best Transcript |
|---|---|---|---|---|
| TALAIAGTLKFNPQT | HLA-DRB1*11:01 | 407.87 | 21 | ENST00000216254.4 |
| GWEFLAFTRLTSELN | HLA-DRB1*11:01 | 26.34 | 1.1 | ENST00000395842.2 |
| VSLFGALVRSRTYDM | HLA-DRB1*11:01 | 9.99 | 0.15 | ENST00000248846.5 |
| SGEKAGCPWSGTGQH | HLA-DRB1*11:01 | 17884.961 | 97 | ENST00000406386.3 |
| SAAAAAAAVIPTVST | HLA-DRB1*11:01 | 7294.73 | 82 | ENST00000381793.2 |
| LASLMVCELAGNLFN | HLA-DRB1*11:01 | 2030.6 | 51 | ENST00000402918.2 |
| AILSHAAREERTEASGY | HLA-DRB1*11:01 | 4605.48 | 70 | ENST00000395741.3 |
| NNKSVNEALNNLLTE | HLA-DRB1*11:01 | 3728.89 | 66 | ENST00000263200.10 |
| QYHVASVLCQRAKVEMS | HLA-DRB1*11:01 | 236.91 | 11 | ENST00000216254.4 |
| SRTLLLAAAGGSSLQ | HLA-DRB1*11:01 | 231.28 | 14 | ENST00000458694.1 |
| KRRSRAGRSGGSHVWTH | HLA-DRB1*11:01 | 6702.69 | 79 | ENST00000262608.8 |
| KTTVVVTTQKRNSRR | HLA-DRB1*11:01 | 70.7 | 4.1 | ENST00000262794.5 |

**Supplementary Figure 2**. Example of color-coded excel peptide sheet from ‘pvacseq create_peptide_ordering_form’. Figure shows an excel sheet with mutated sequence containing 25 flanking amino acids on either side, with mutated amino acid(s) underlined, representative MHC class I binder bolded, representative MHC class II binder colored in red, along with best class I/II binders-HLA information and the corresponding molecular weight. User-defined problematic amino acids (eg. Cysteine) are printed in larger font.

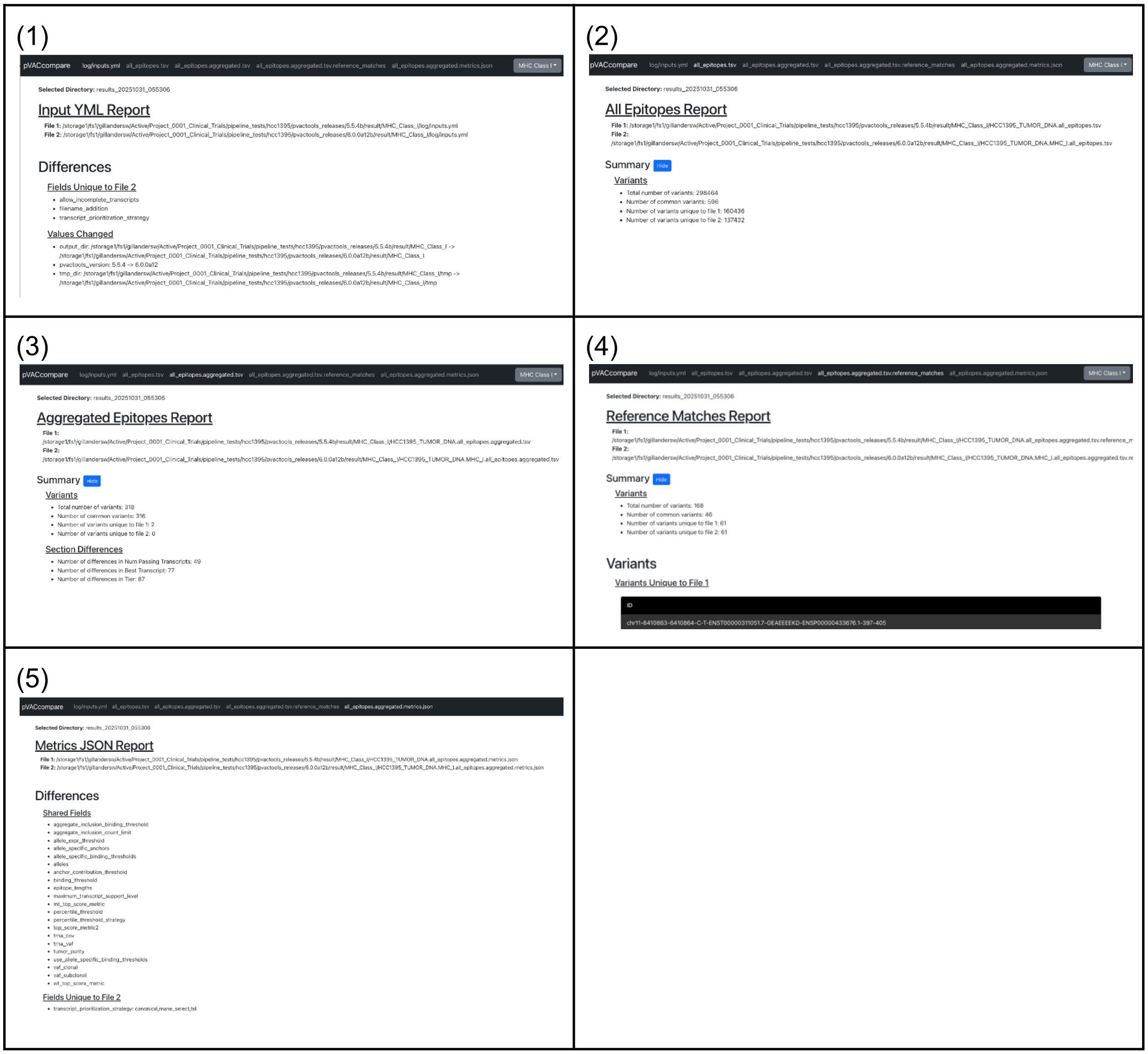

**Supplementary Figure 3**. pVACcompare has 5 output files.
(1) log/inputs.yml (which specifies differences in input file and pVAC run set up);
(2) all_epitopes.tsv (summarizes total common and unique variants);
(3) all_epitopes.aggregated.tsv (difference in variants, best representative transcript and candidate Tier);
(4) all_epitopes.aggregated.tsv.reference_matches (summarizes count and identify of variants with reference proteome match across runs);
(5) all_epitopes.aggregated.metrics.json (summarizes the neoantigen metrics, or setup in each run, facilitating comparison of runs from different pVAC versions).

A

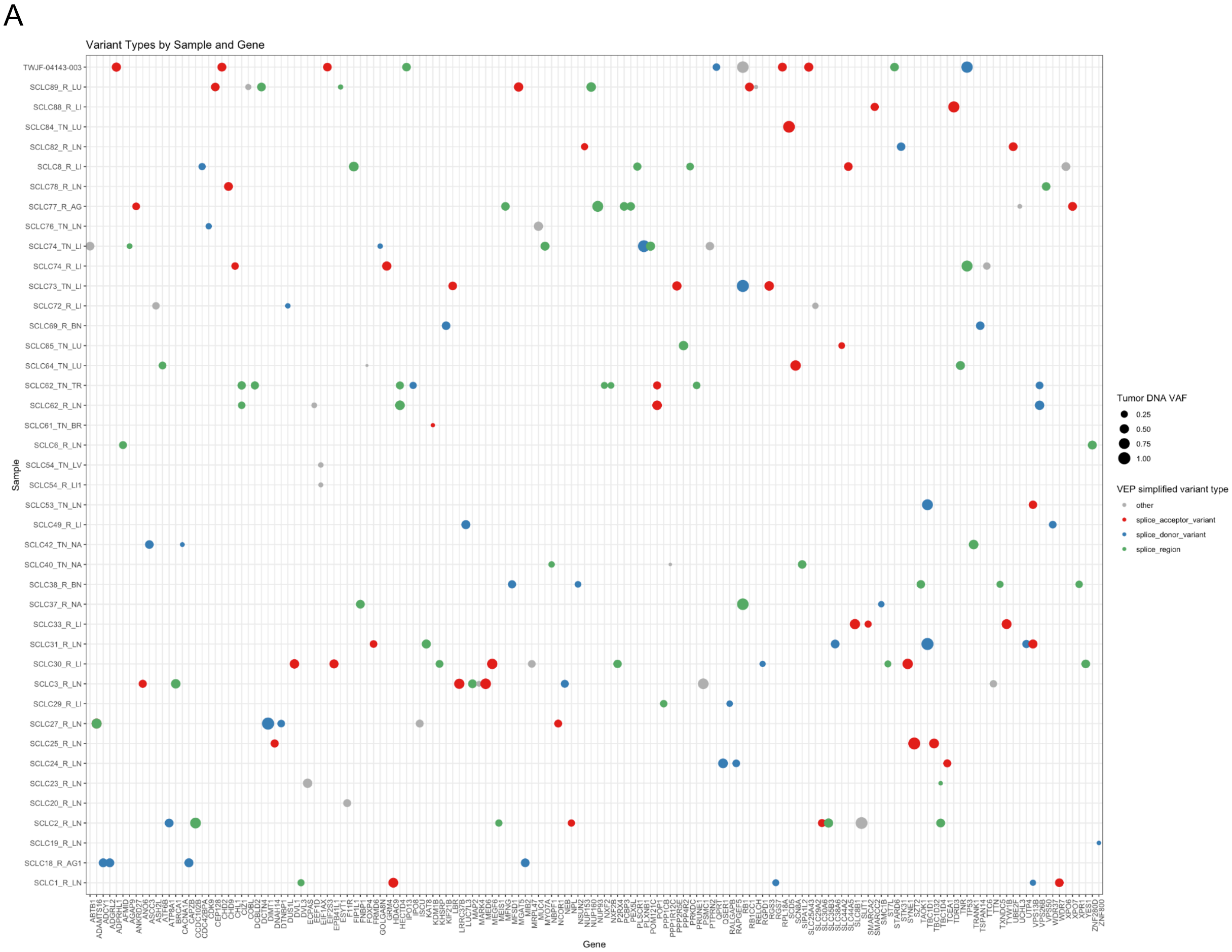

Variant Types by Sample and Gene
Sample
Gene
Tumor DNA VAF
0.25
0.50
0.75
1.00
VEP simplified variant type
other
splice_acceptor_variant
splice_donor_variant
splice_region

B

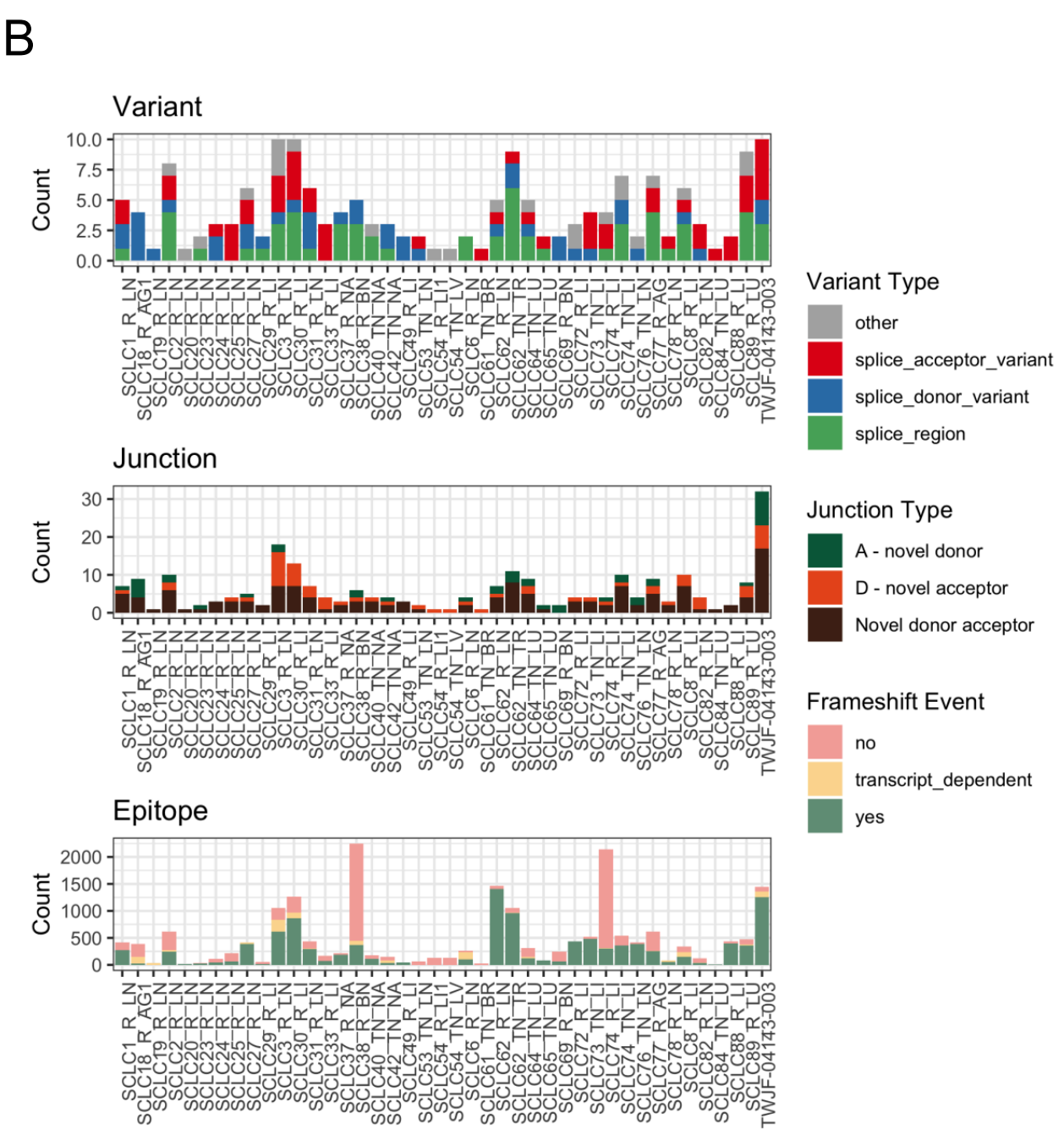

Variant
Count
Variant Type
other
splice_acceptor_variant
splice_donor_variant
splice_region
Junction
Junction Type
A - novel donor
D - novel acceptor
Novel donor acceptor
Frameshift Event
no
transcript_dependent
yes
Epitope

**Supplementary Figure 4**. **pVACsplice predicts multiple cis-splicing variants in SCLC cohort**. **(A)** Dot plot summarizes the cis-splicing variants observed across the cohort. Dots are sized according to variant tumor DNA VAF, and colored based on VEP annotation. Putative neoantigens are identified for key drivers TP53 and RB1 - the two most frequently altered tumor suppressor genes in SCLC.**(B)**: Summary of pVACsplice outputs including splice variants, junctions, and predicted MHC binders across the cohort. Variant counts in this panel refer to the number of variants passing the pVACsplice default filters. Junction counts illustrated here are for junctions associated with these variants that are confirmed as absent from normal tissues in GTEx.

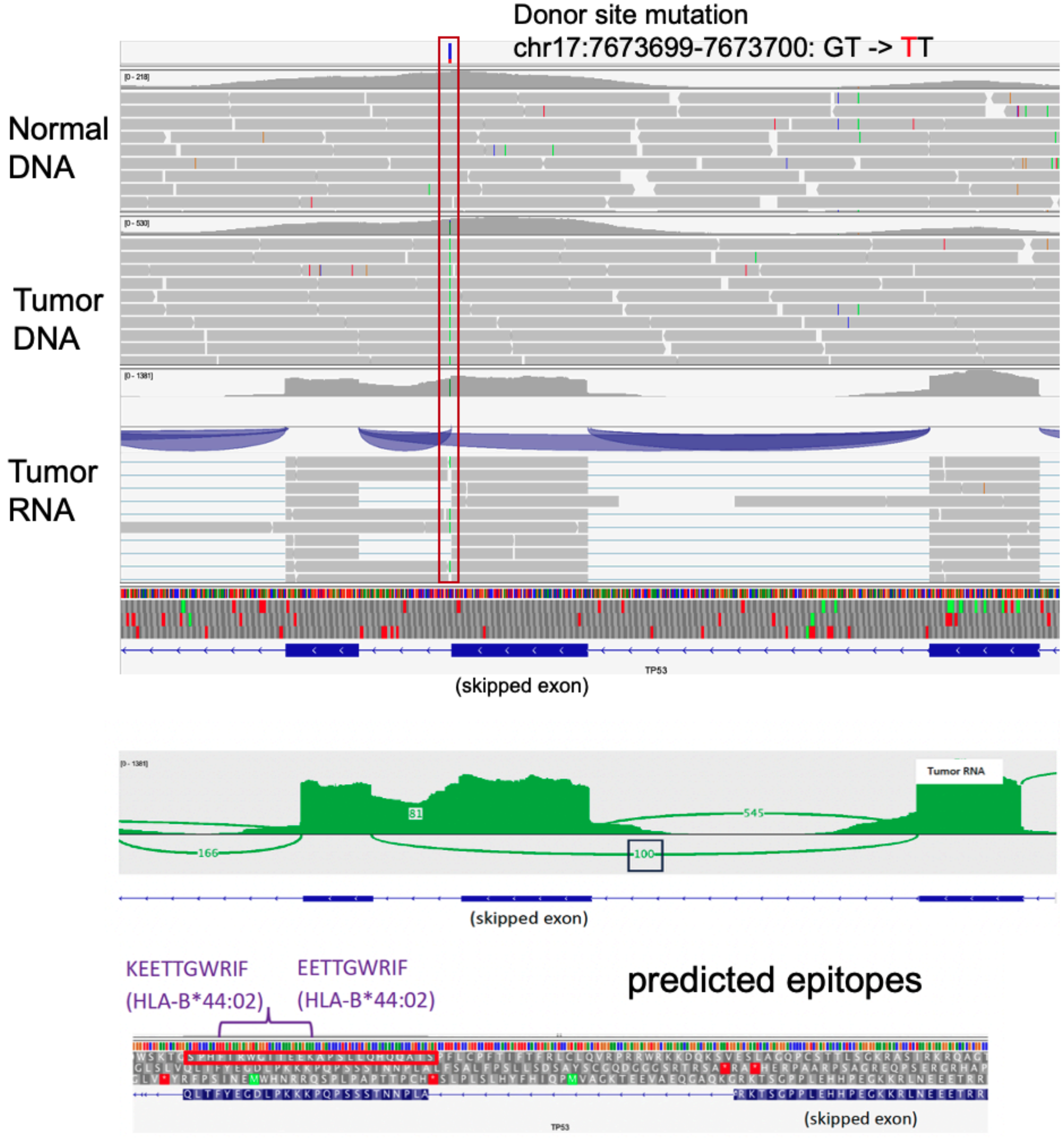

**Supplementary Figure 5**. Illustrative example: a mutation at the splicing donor site causes an exon skipping event and creates multiple potential epitopes as predicted by pVACsplice. The donor site mutation is tumor-specific, disrupting the canonical splicing donor dinucleotide GT, causing skipping of exon 4, thereby creating a novel exon3-exon5 junction that is robustly supported (with 100 reads assigned to the novel junction). This event is predicted to produce a frameshift and shortened TP53 protein. The mutant protein sequence contains predicted strong MHC binders for the patient's HLA allele.

ETVMPSALVRGPLQPVSLLACPFRHRC
GPGPG
RHEMVAKPPAMCSRFAKDLRPEQYIKNSFQ
GGS
GYIRQVGDFHQVIIRGRGHILPYDQPLRAFDMI
GGS
ALEREEEEERERARLWERRQQRKNREAFQTF
HHHH
PNQTVVKMFLVTFDFSNMPAAHMTFLRHRLFL
GGS
SGGGPGAQHSAMPAKSKEELSEGSRKKKT
GGS
CEVGMAYSMEKYRRTDNMARVMVRYMKY

pvacV4_IC1500_Clip5

SCEVGMAYSMEKYRRTDNMARVMVRYMKYLRQK
GGS
ALEREEEEERERARLWERRQQRKNREAFQTF
NETVMPSALVRGPLQPVSLLACPFRHRC
GPGPG
GYIRQVGDFHQVIIRGRGHILPYDQPLRAFDMI
GGS
NMKRHEMVAKPPAMCSRFAKDLRPEQYIKNSFQ
HHHH
PNQTVVKMFLVTFDFSNMPAAHMTFLRHRLFLV
GGS
SGGGPGAQHSAMPAKSKEELSEGSRKKKTKKAD

pvacV5_IC1500_Clip5

**Supplementary Figure 6**. Vector designs for test case G104 under the IC1500_Clip5 setting, shown side-by-side for pVACvector version 4 and version 5. Version 5 produces a solution that fully preserves the binding core (GSRKKKTKK). Binding cores are colored red. Locations of binding core with clipping event of interest are highlighted in yellow.

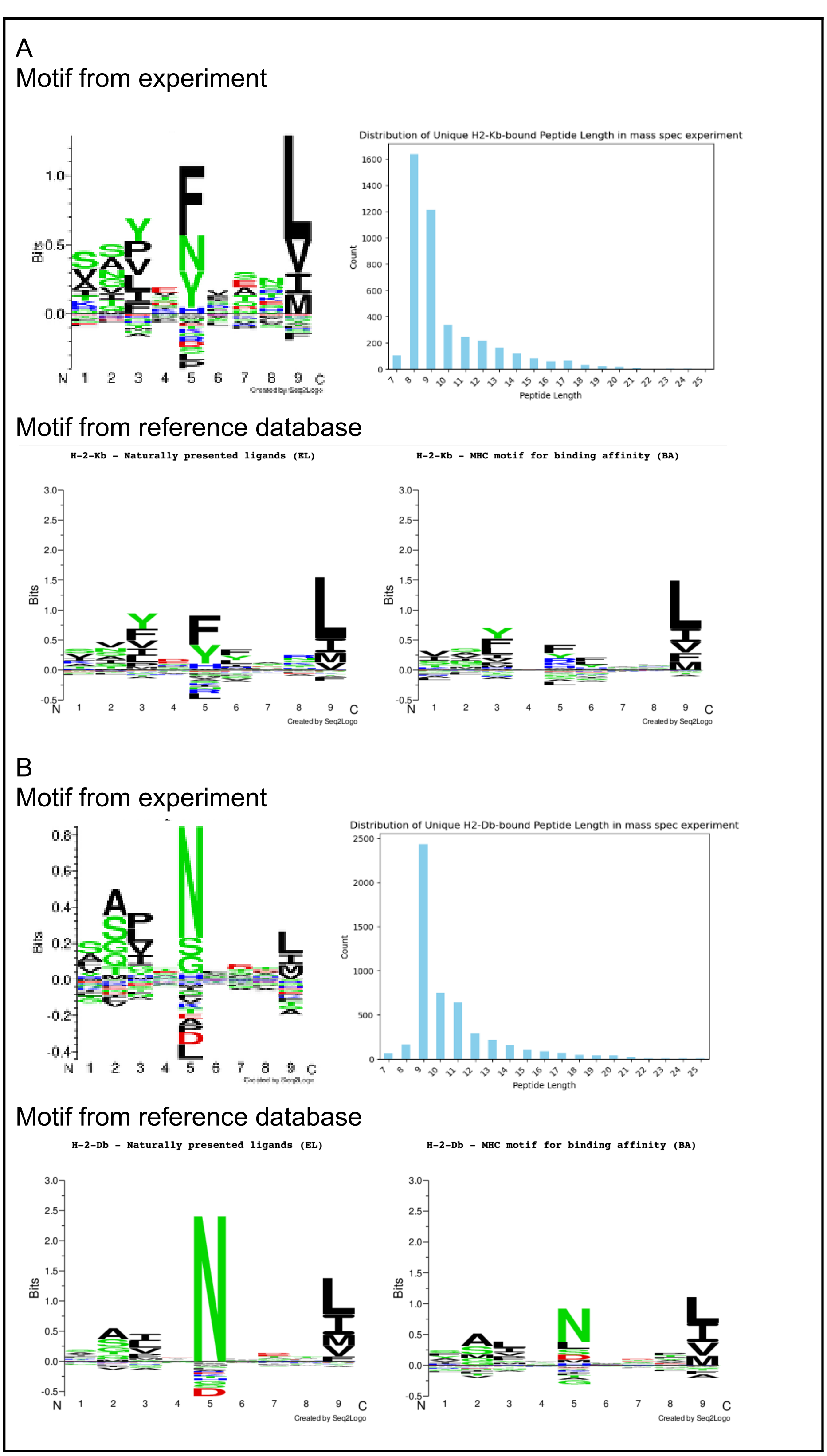

**Suppl Fig 7**. Mass-spec identified MHC binders from experiment have sequence motifs similar to motifs from reference database NetH2pan. (A) H-2-Kb binders. (B). H-2-Db binders.

# Supplementary tables

**Sup Table 1**. Tiers in pVACtools. Tier available in each tool is marked with y (yes).

| Tier | pVACseq | pVACsplice | pVACfuse |
|---|---|---|---|
| Pass | y | y | y |
| Poor Binder | y | y | y |
| Ref Match | y | y | y |
| Poor Transcript | y | y | |
| Low Expr | y | y | y |
| Anchor | y | | |
| Subclonal | y | y | |
| Prob Pos | y | y | y |
| Poor | y | y | y |
| No Expr | y | y | |
| Low read support | | | y |

**Sup Table 2**. Summary of successful case count and run time (average and median) in benchmark

| pvacVer | RunParameter | Successful design | avg_runtime | median_runtime |
|---|---|---|---|---|
| pvacV4 | default | 11 | 70.9 | 52.1 |
| pvacV5 | default | 11 | 57.5 | 43 |
| pvacV4 | IC1500 | 8 | 334.6 | 201.9 |
| pvacV5 | IC1500 | 8 | 131.2 | d97.2 |
| pvacV4 | IC1500_Clip5 | 9 | 326 | 257.1 |
| pvacV5 | IC1500_Clip5 | 9 | 134.1 | 117 |

# Author contributions

M.H.H. and S.K. wrote the manuscript and produced the visualizations. M.G. and O.L.G. conceived and developed the protocol. M.R., L.H., H.X., E.S., K.C.C., C.F.L., Z.S., K.S., J.X.Y., J.L., M.K. and I.R. analyzed and interpreted data. M.H.H., S.K., C.A.M, J.W., R.G., G.D., T.M.J., S.P.G., T.A.F., R.K., R.D.S., W.G., O.L.G and M.G. revised the manuscript. O.L.G. and M.G. participated in project administration and supervision, and provided resources. R.G., J.W, and W.G. provided clinical oversight.

# Acknowledgements

We are grateful to the patients and their families for the donation of their samples and participation in clinical trials. M.G., O.L.G., R.K. and S.K. were supported by NIH National Cancer Institute (NCI) under award number R01CA301054. M.G. and O.L.G. were supported by the NIH National Cancer Institute under award numbers U01CA209936, U01CA231844, and U24CA237719. M.G. was supported by the V Foundation for Cancer Research under award number V2018-007. M.G., O.L.G., H.X., W.E.G., T.A.F were supported by the NCI under award number U01CA248235. W.E.G, T.A.F were supported by the Siteman Cancer Center (P30CA91842). T.A.F, R.D.S, O.L,G, M.G were supported by LLS SCOR. T.A.F was also supported by Lymphoma Research Foundation, Blood Cancer United, the Howard Steinberg Family Foundation, and the Paula and Rodger Riney Blood Cancer Initiative. M.G., O.L.G. and W.E.G. were supported by the NCI/Leidos Biomedical Research Subcontract 20X012F5P50. S.P.G. was supported by the Foundation for Barnes-Jewish Hospital. K.S. was supported by NIH/NIGMS T32-GM139774. W.E.G was supported by CA/NCI NIH T32-CA009621, Susan G. Komen for the Cure (KG111025), B101/Centene Corporation P19-00559, Alvin J. Siteman Cancer Center Investment Program grant 4035, CA/NCI NIH R01-CA240983, MedImmune AstraZeneca 004135-946897, P50 CA196510, and P50 CA272213. This work was also supported in part by the Washington University Institute of Clinical and Translational Sciences from the National Center for Advancing Translational Sciences (NCATS) of NIH under award number UL1TR002345. Finally, this work was supported by a gift from the Goldberg Family Foundation.

# Author Disclosures

K.S., E.S., K.C.C, M.G., and O.L.G. are consultants for the Jaime Leandro Foundation and Pathfinder Oncology. TAF holds equity in Wugen, Orca Bio, and Indapta Therapeutics; and has research funding from ImmunityBio, AI Proteins. R.K. receives royalty distributions through Johns Hopkins Technology Ventures from licensing agreements with Exact Sciences Corp. (formerly Thrive Earlier Detection Corp.), Genentech Corp., Agios Pharmaceuticals, Inc., and Servier Pharmaceuticals;